\documentstyle[preprint,eqsecnum,aps]{revtex}
\draft  

\begin{document}

\title{Completeness Rules for Spin Observables\\
           in Pseudoscalar Meson Photoproduction}
\author{Wen-Tai Chiang~\cite{byline} and Frank Tabakin~\cite{byline}}
\address{Department of Physics and Astronomy, University of Pittsburgh,
         Pittsburgh, PA~15260}
\date{\today}
\maketitle

\begin{abstract}

The number and type of measurements needed to ascertain the amplitudes
for pseudoscalar meson photoproduction are analyzed in this paper. 
It is found that 8 carefully selected measurements can determine the four
transversity amplitudes without discrete ambiguities.
That number of measurements is one less than previously believed. 
We approach this problem in two distinct ways: (1) solving for the
amplitude magnitudes and  phases directly; and (2) using a bilinear
helicity product formulation to map an algebra of measurements over to
the well-known algebra of the $4\times 4$ Gamma matrices.
It is shown that the latter method leads to an alternate proof that
8 carefully chosen experiments suffice for determining the transversity
amplitudes completely.
In addition, Fierz transformations of the Gamma matrices are used to
develop useful linear and nonlinear relationships between the spin
observables.
These  relationships not only help in finding complete sets of experiments,
but also yield important constraints between the 16 observables for this
reaction.

\end{abstract}


\section{Introduction}
Interest in the photoproduction of pseudoscalar mesons has been revived
 now that experiments of unprecedented precision are imminent.
With the development of new electron accelerator facilities (such as TJNAF)
 along with both polarized beams and targets and with the CLAS detector,
 it will soon be possible to measure various spin observables with precision.
These observables include the differential cross section, $\sigma(\theta)$,
 plus three single spin observables ($\Sigma$, $T$ and $P$),  which we
 denote as Type $\cal S$ measurements.
In addition, there are twelve double spin observables which can be
 classified into three types: beam-target ($\cal BT$), beam-recoil
 ($\cal BR$) and  target-recoil ($\cal TR$) spin observables.
The classic Barker, Donnachie and Storrow (BDS)~\cite{BDS} paper is one 
 of the standard references on how to select measurements
 to fully determine the four (complex) total pseudoscalar
 meson photoproduction amplitudes.
In this paper, we also address that question. 

It is well-known that, without considering discrete ambiguities,  
 seven measurement are needed to determine the four
 helicity amplitudes (four magnitudes plus three phases)
 up to an arbitrary overall phase.
However,  it is necessary to resolve all discrete ambiguities  
 to  extract complete 
 information from experiments. 
In  BDS,  the following rule~\cite{BDS} (herein called the BDS rule)
 was promulgated:

\begin{quote}
{\it In order to determine all amplitudes without discrete ambiguities, 
one has to measure
      \underline{five} double spin observables along with the four type
      $\cal S$ measurements, provided  no four double spin observables
are selected from the
      same set of $\cal BT$, $\cal BR$ and $\cal TR$.}
\end{quote} Thus, they say nine experiments are required.

Recently, Keaton and Workman (KW)~\cite{Workman} argued that 
 selecting a complete set of 
 observables is more complicated than the  above BDS rule.  
However, KW were not able to provide sufficient conditions for resolving
 all ambiguities.
Their work inspired us to investigate the problem of determining
which experiments can provide a ``complete set," e.g. those
 experiments which suffice
to determine the basic amplitudes free of discrete or continuous
ambiguities.
Here we  confirm the KW result~\cite{Workman} that there are cases
 obeying the BDS rule that still leave  unresolved ambiguities.
To our surprise, we also find that  \underline{four} appropriately chosen
 double spin observables, along with the four Type $\cal S$ measurements
 suffice to resolve all ambiguities. This is our major result.
It is illustrated first by using the explicit approach used in BDS;
 namely,
by solving for the magnitudes and phases of  
transversity helicity amplitudes.  Transversity amplitudes provide the
advantage of having all Type ${\cal S}$ (single spin) observables expressed
in terms of the amplitude magnitudes only.  The double spin observables
are then needed to determine the phases of the transversity amplitudes. 

Another approach is also provided in this paper.
In this alternate approach, Hermitian versions of the
usual $4 \times 4$ Gamma matrices are used to express 
  all observables as bilinear products of
helicity amplitudes.  In that way, algebraic relations between
observables (an algebra of measurements) are mapped into the
well-known algebra of the $4 \times 4$ Gamma matrices.       
For example,  important relationships between spin observables
 are derived here by applying the Fierz identities
 to products of Gamma matrices.  This procedure, as explained later,  
yields useful relationships between observables which serve  
to select complete sets of observables.
One benefit of this bilinear helicity product (BHP) approach is 
that it can be generalized to
 other reactions~\cite{vmeson}.

In Section~\ref{sec:form}, we present the bilinear helicity
 product analysis of spin observables.
In Section~\ref{sec:amb}, we give a general discussion of the discrete
 ambiguities, with emphasis on linear and nonlinear ambiguities.
In Section~\ref{sec:comp set}, we give an example of a complete set of eight
 measurements which resolve all ambiguities, and then present Tables of
 all such sets of observables.
In Section~\ref{sec:fierz}, relations among spin observables are derived
 using the Fierz identities, which are then used to confirm the
complete sets of observables deduced earlier.

We assume that the four Type $\cal S$ observables
 are always measured and that the problem is to select the
 double spin observables which will yield unambiguous total
amplitudes.  We do not deal with the problem of extracting
partial wave amplitudes.

\section{Bilinear Helicity Products}
\label{sec:form}
In this section, we introduce the bilinear helicity product (BHP)
 formulation for discussing spin observables, following the conventions in
 references~\cite{vmeson,psmeson}.

The pseudoscalar meson photoproduction reaction is completely described by
 four complex helicity amplitudes: $H_1$, $H_2$, $H_3$ and $H_4$,\footnote
 {These are also often denoted by $S_1$, $N$, $D$ and $S_2$, where $S$
  refers to single-flip, $D$ double-flip and $N$ no-flip amplitudes.}
 at each energy and angle.
The 16 spin observables, $\Omega^{\alpha},$ consist of the differential cross
 section $\sigma(\theta)$, plus 3 single and 12 double spin observables.
 Expressions for these 16 observables
 in terms of the helicity,
 transversity and BHP forms are presented
in Table~\ref{tbl:spinob}.\footnote{ For convenience, some of the  
 $\Omega^{\alpha}$'s in Table~\ref{tbl:spinob} were defined with different
   signs than in Ref.\cite{vmeson}.}  
All of the 16 observables can  be expressed in bilinear helicity product
 (BHP) form~\cite{vmeson}:
\begin{equation}
  \check{\Omega}^{\alpha} = \Omega^{\alpha}\, {\cal I}(\theta)
  = \frac{1}{2}\,H^{\ast}_i\,\Gamma^{\alpha}_{ij}\,H_j
  \equiv \frac{1}{2}\langle H|\Gamma^{\alpha}|H\rangle ~,
  \qquad \alpha = 1, \cdots 16 ~,
\end{equation}
 where summation over repeated indices is implied.
We define ${\cal I}(\theta) = (k/q)\sigma(\theta)$, where $k$ and $q$ are
 the momenta of the initial and final states in the center-of-mass frame.
The $\check{\Omega}^{\alpha}$ are the
 ``profile function"~\cite{vmeson,psmeson} forms of the spin
 observables, and $\Gamma^{\alpha}$ matrices are the sixteen $4\times4$
 Hermitian Gamma matrices.
See Appendix~\ref{ap:gamma} for details about the $\Gamma^{\alpha}$ matrices.
The 16 spin observables are also classified in Table~\ref{tbl:spinob},
 as four sets: ${\cal S}$,
 ${\cal BT}$, ${\cal BR}$ and ${\cal TR},$ with four observables in
each set.

A unitary transformation $U^{(4)}$ acting on both the helicity amplitudes
 and the $\Gamma^{\alpha}$ matrices:
\begin{eqnarray}
  H_i &\longrightarrow& b_i = U^{(4)}_{ij} H_j \\
  \Gamma^{\alpha} &\longrightarrow&
   \widetilde{\Gamma}^{\alpha} = U^{(4)}\Gamma^{\alpha} U^{\dag(4)}~,
\end{eqnarray}offers a means of
 altering  amplitudes without changing observables.  
Such changes in description without changing observables 
are called canonical transformations, as in mechanics.
A particularly useful unitary transformation of this type is the transversity
 choice~\cite{vmeson},
\begin{equation}
  U^{(4)} = \frac{1}{2}
  \left(\begin{array}{rrrr}
    1 & -i &  i &  1 \\
    1 &  i & -i &  1 \\
    1 &  i &  i & -1 \\
    1 & -i & -i & -1
  \end{array} \right) ~,
\label{eq:u4}
\end{equation}   
 which involves rotating the helicity quantization axis to the direction
 normal to the scattering plane. 
The sixteen spin observables can be expressed in this {\it transversity
 basis} by
\begin{equation}
  \check{\Omega}^{\alpha} = \Omega^{\alpha}\, {\cal I}(\theta)
  = \frac{1}{2}\,b^{\ast}_i\,\widetilde{\Gamma}^{\alpha}_{ij}\,b_j
  = \frac{1}{2}\langle b|\widetilde{\Gamma}^{\alpha}|b\rangle ~,
  \qquad \alpha = 1,  \cdots 16 ~.
\end{equation}
Note that the corresponding $\widetilde{\Gamma}^{\alpha}$ matrices for the
 four Type ${\cal S}$ measurements are diagonal in the transversity basis,
 i.e., these observables involve combinations of the  squared magnitudes,
 $\pm |b_i|^2,$ of the transversity helicity amplitudes.
If all four of the Type ${\cal S}$ observables
 are measured (as assumed in this paper), then the double
 spin observables are used to determine only the phases of the transversity
 amplitudes.
The explicit forms of the Gamma matrices in the transversity basis,
 $\widetilde{\Gamma}^{\alpha}$, are presented in Appendix~\ref{ap:gamma}. 
In this paper, we will work mainly in the transversity basis.

After the above transversity
transformation,  both the amplitudes and the Gamma matrices are
changed, without altering the observables. In the next section, we
will introduce unitary transformations corresponding to
discrete changes of the amplitudes that can change observables. 

This BHP form will be used first to discuss such
  discrete ambiguities and later
for a general approach to the completeness problem.

\section{Discrete Ambiguities}
\label{sec:amb}

\subsection{General definition of discrete ambiguities}

The extraction of reaction amplitudes from experiments poses an interesting,
 and sometimes difficult task, because it is a nonlinear problem.
To gain insight into the general nature of this problem and to define
discrete ambiguities broadly, let us consider a
 reaction described by $N$ complex amplitudes.  For
pseudoscalar meson photoproduction $N=4$, and
we deal with a $4\times 4$ Gamma algebra.  For the
general $N$ case one also has a BHP form,  but
it is represented by a
$ N \times N$ Clifford algebra.  There are $N^2$
 {\it linearly independent} experimental observables which are linear
 combinations of the $N^2$ bilinear products of the $N$ amplitudes.
One might assume that $2N-1$ appropriately chosen observables can determine
 these $N$ amplitudes, apart from an overall phase factor.
However, these $N^2$ observables are {\it nonlinearly dependent} on
 each other, and several discrete solutions may satisfy these $2N-1$
 measurements simultaneously.
Therefore, more than $2N-1$ experiments are needed to resolve ambiguities,
 and the number of additional measurements required is not a fixed number,
 but depends on the type of measurements already performed~\cite{Morav}.
Here we study these discrete ambiguities, following 
 Dean and Lee~\cite{DL}, and find ways to resolve them.
Some of the following discussion is equivalent to the methods
 proposed by Keaton and Workman~\cite{Workman}.

In general, the observables (as profile functions)
 $\{{\check{\Omega}}^{\alpha}\}$ can  be
 expressed in a bilinear product form with the $N$ helicity amplitudes
 $H_1 \cdots H_N$:
\begin{equation}
  {\check{\Omega}}^{\alpha}= {\Omega}^{\alpha} {\cal I}(\theta)
 =\frac{1}{2} H^{\ast}_i \Gamma^{\alpha}_{ij} H_j ~,
\end{equation}
 where $\Gamma^{\alpha}$ are Hermitian $N \times N$ matrices.
An ambiguity occurs in extracting the $N$ amplitudes $H_i$ from a
  {\it subset of measurements}  $\{{\Omega}^{\alpha}\}\equiv 
{\Omega}^{\alpha_1}\cdots{\Omega}^{\alpha_M},$ where $M < N \times N, $ 
 when there exists a transformation
 on the amplitudes $H_i$ under which that subset of 
 observables $\{{\Omega}^{\alpha}\}$ is invariant.
To remove that ambiguity,  one need to enlarge the
 subset ${M\rightarrow M+1}$ wisely, until no such transformation exists.
That defines the process for removing discrete ambiguities.

A trivial case  of an ambiguity is an overall phase transformation applied
to all $N$ amplitudes $H_i \rightarrow e^{i\delta} H_i$ with the real
 $\delta$ independent of $i$.
Since $M = N^2$,  there is no way to
remove this ambiguity,  which shows that only the relative
 phases of the amplitudes can be determined.
Other nontrivial ambiguities will be discussed later.
If the set  $\{{\Omega}^{\alpha}\}$  of $M$ observables 
is sufficient to eliminate all 
 ambiguities, then a unique set of amplitudes can be extracted.
In that case, we call $\{{\Omega}^{\alpha}\}$ a {\it complete set} of
 measurements.
Linear and nonlinear transformations of  
the amplitudes can be defined to perform the above test.

As an example of a linear type of transformation that could leave 
a subset of observables unchanged,
 consider the following unitary transformation applied to all $N$
helicity amplitudes:
\begin{equation}
  H_i \longrightarrow H'_i = L_{ij} H_j ~,
\end{equation}
where $L$ is chosen unitary to conserve the differential cross section
 ${\cal I}(\theta) =\frac{1}{2} H^{\ast}_i \delta_{ij} H_j =
\frac{1}{2} \sum_i |H_i|^2$.~\footnote{
Here we suppress density of states factors and use the fact that 
the cross section is the sum of magnitude-squared helicity amplitudes}
If there exists a unitary $L$ commuting with all $\Gamma^{\alpha}$'s
in the $M<N^2$ subset ${\Omega}^{\alpha_1}\cdots {\Omega}^{\alpha_M} $, 
 i.e.,
\begin{equation}
  L^{\dag} \Gamma^{\alpha_n} L = \Gamma^{\alpha_n} \qquad n= 1 \cdots M<N^2,
\label{eq:lin}
\end{equation}
 then for members of that subset
\begin{equation}
  {\check{\Omega}}^{\alpha_n}
   = \frac{1}{2} H^{\ast}_i \Gamma^{\alpha_n}_{ij} H_j
   = \frac{1}{2} H^{\ast}_i (L^{\dag}\Gamma^{\alpha_n}L)_{ij} H_j
   = \frac{1}{2} H^{\prime\ast}_i \Gamma^{\alpha_n}_{ij} H'_j ,
\end{equation}which shows that the subset of observables
$\{{\Omega}^{\alpha}\}$ are invariant
under $L$ and 
  can not be used to distinguish
 between amplitudes $H_i$ and $H'_i$.
Then, there is a {\it linear ambiguity}.

Next let us now consider an antilinear transformation
 acting on all $N$ helicity amplitudes:
\begin{equation}
  H_i \longrightarrow H'_i = A_{ij} H^{\ast}_j ~,
\end{equation}
 where $A$ is unitary.
Any $A$ satisfying
\begin{equation}
  (A^{\dag} \Gamma^{\alpha_n} A)^T = \Gamma^{\alpha_n},
\label{eq:antilin}
\end{equation} where $\alpha_n$ corresponds to any
observable in the  $M<N^2$
 subset ${\Omega}^{\alpha_1}\cdots {\Omega}^{\alpha_M}, $   
 defines an {\it antilinear ambiguity} for $\{{\Omega}^{\alpha_n}\}$ because
\begin{eqnarray}
{\check{\Omega}}^{\alpha_n}
  &=& \frac{1}{2} H^{\ast}_i \Gamma^{\alpha_n}_{ij} H_j \nonumber \\
  &=& \frac{1}{2} H^{\ast}_i (A^{\dag}\Gamma^{\alpha_n}A)_{ji} H_j\nonumber\\
  &=& \frac{1}{2} H^{\ast}_i A^{\dag}_{jk} \Gamma^{\alpha_n}_{kl} A_{li} H_j
      \nonumber \\
  &=& \frac{1}{2} (A_{li}H^{\ast}_i)\Gamma^{\alpha_n}_{kl}(A^{\dag}_{jk} H_j)
      \nonumber \\
  &=& \frac{1}{2} H^{\prime\ast}_k \Gamma^{\alpha_n}_{kl} H'_l ~,
\end{eqnarray}
 which shows that
members of  the measurement subset $ {\Omega}^{\alpha_n}$ can not
 distinguish between amplitudes $H_i$ and $H'_i.$

\subsection{Discrete ambiguities for pseudoscalar meson photoproduction}

For pseudoscalar meson photoproduction ($N=4$), we have expressed
 the sixteen spin observables in BHP form using the transversity basis,
\begin{equation}
  \check{\Omega}^{\alpha}
   = \frac{1}{2}\,b^{\ast}_i\,\widetilde{\Gamma}^{\alpha}_{ij}\,b_j ~.
\end{equation}
To find associated discrete ambiguities, we need to look for matrices $L$ and
 $A$ which satisfy Eq.~(\ref{eq:lin}) and (\ref{eq:antilin}), respectively.
Since the sixteen Hermitian $\widetilde{\Gamma}^{\alpha}$ matrices form
 a basis for $4\times4$ matrices, it is sufficient to find
 $\widetilde{\Gamma}^{\alpha}$ matrices satisfying Eq.~(\ref{eq:lin})
 and (\ref{eq:antilin}).
Suppose that we always measure the four Type ${\cal S}$ observables:
 $\check{\Omega}^1$, $\check{\Omega}^4$, $\check{\Omega}^{10}$,
 $\check{\Omega}^{12}.$ 
 The only $\widetilde{\Gamma}^{\alpha}$ matrices commuting with all four of
 those $\widetilde{\Gamma}^{\alpha}$ matrices in Type ${\cal S}$ are:
  \begin{equation}
  L = \widetilde{\Gamma}^4, \widetilde{\Gamma}^{10}, \widetilde{\Gamma}^{12}
      \qquad \mbox{for} \ {\cal S}
\end{equation}
 (where $L = \widetilde{\Gamma}^1$ is not listed because it obviously 
leaves all amplitudes
 unchanged).  Those $\widetilde{\Gamma}^{\alpha}$ matrices satisfying 
the antilinear transformation case, Eq.~(\ref{eq:antilin}), are:
 \begin{equation}
  A = \widetilde{\Gamma}^6, \widetilde{\Gamma}^8, \widetilde{\Gamma}^{13},
      \widetilde{\Gamma}^{15} \qquad
      \mbox{for} \ {\cal S}.
\end{equation}
So $L = \{\widetilde{\Gamma}^4, \widetilde{\Gamma}^{10},
 \widetilde{\Gamma}^{12}\}$ are possible candidates for testing for
 linear ambiguities in any subset of measurements which includes
 type ${\cal S}$ measurements.
Similarly, $A = \{\widetilde{\Gamma}^6,\widetilde{\Gamma}^8,
 \widetilde{\Gamma}^{13},\widetilde{\Gamma}^{15}\}$ test for
 antilinear ambiguities.
We believe that all other transformations, assuming type ${\cal S}$
 measurements, can be constructed from the above basic linear and antilinear
 unitary transformations. 

All sixteen spin observables remain either unchanged or simply
 change sign under these  basic linear and antilinear transformations.
The results are given in Table~{\ref{tbl:amb} (some of these cases
 are in Ref.~\cite{Workman}).
We are working with transversity amplitudes and correspondingly with the
 transformed matrices $\widetilde{\Gamma}$.
The parallel results are expressed
in the helicity basis in Appendix~\ref{ap:kw},  wherein the connection
to the results of Ref.~\cite{Workman} is made. 

If a subset of measured observables are invariant under one of these
 linear or antilinear transformations, then a discrete ambiguity exists.
For example, if we measure $G, F, O_z, C_x, T_x$ and $L_z ,$ in addition to
 type ${\cal S}$, since they are all unchanged under the antilinear
 transformation with $A = \widetilde{\Gamma}^6$, these 4 + 6 = 10 spin
 observables cannot resolve all ambiguities.\footnote{In this case, the
 transformation is $b_1 \leftrightarrow - b^*_2$ and
 $b_3 \leftrightarrow b^*_4,$  see Eq.~(A5). }
Note that the BDS rule is violated in this case.
Therefore, {\it to determine the amplitudes uniquely, one has to choose
 a set of spin observables that are not all invariant under these
 $L$ and $A$ transformations}.
Unfortunately, the above statement provides only {\it necessary} but not 
 {\it sufficient} conditions to determine unique solutions, since there are
 also {\it nonlinear} ambiguities which are relatively difficult to resolve.

To clarify the above discussion,  we note that some transformations 
of the basic amplitudes leave some set of observables
unchanged,  while other observables simply change sign.  For example,
the replacement $b_3 \rightarrow -b_3,$ and
$b_4 \rightarrow -b_4,$  leave  the eight observables
$\Omega_{1,4,10,12}$ (Type ${\cal S})$ and the $\Omega_{6,13,8,15}$
(Type ${\cal TR})$ 
 unchanged, while the sign of the eight observables
$\Omega_{3,5,9,11}$ (Type ${\cal BT})$  and 
 $\Omega_{14,7,16,2}$ (Type ${\cal BR})$ are changed, see Table I. 
If none of these sign changed observables are among those measured, then
we have an ambiguity in determining $b_3\ \&\ b_4.$  This particular
transformation of the amplitudes can be represented 
as $b_i' = U_{ij}b_i $
with:
\begin{equation}
  U  =  
  \left(\begin{array}{rrrr}
    1 & 0 & 0 & 0 \\
    0 & 1 & 0 & 0 \\
    0 & 0 &-1 & 0 \\
    0 & 0 & 0 &-1
  \end{array} \right) ~,
\label{eq:udisc1}
\end{equation} which is identical to $\widetilde{\Gamma}^4$.
Now consider the effect of such a transformation on all of the observables
 $\Omega^\alpha$.
We have:
\begin{equation}
  \check{\Omega}^{\alpha} = \Omega^{\alpha}\, {\cal I}(\theta)
  = \frac{1}{2}\,b^{\ast}_i\,\widetilde{\Gamma}^{\alpha}_{ij}\,b_j
  \rightarrow  
\frac{1}{2} b^{\ast}_i\, U^\ast_{ k' i} 
\widetilde{\Gamma}^{\alpha}_{k'k} U_{kj}\, b_j  ~,
  \qquad \alpha = 1,  \cdots 16 ~.
\label{eq:udisc2}
\end{equation}
Since for our particular example $U \rightarrow \widetilde{\Gamma}_4$,
 the effect of this discrete transformation on the transversity amplitudes
  is equivalent to the following substitution:
  
 \begin{equation}
 \widetilde{\Gamma}^{\alpha} \longrightarrow  
  \widetilde{\Gamma}^4\, \widetilde{\Gamma}^{\alpha}\,\widetilde{\Gamma}^4.    
 \end{equation}
The above effect of $\widetilde{\Gamma}^4$ on $\widetilde{\Gamma}^{\alpha}$,
exactly duplicates the sign changes indicated above that are induced
by the $b_{3,4} \rightarrow -b_{3,4}$ substitution. 
 This result is also seen in the third column of Table I and
the first column of Table II . 

We wish to find a subset of measurements that can be used to deduce a unique
 set of transversity amplitudes.  Once accomplished,
the helicity amplitudes can be obtained by the inverse of
Eq.~(\ref{eq:u4}). Two different approaches to this problem
are presented:
 in Section~\ref{sec:comp set}, we solve for the phases of the transversity
 amplitudes directly from spin observables; in Section~\ref{sec:fierz},
 we derive relations between spin observables from the Fierz identities
 of the $\widetilde{\Gamma}^{\alpha}$ matrices.

\section{Complete Set of Measurements}
\label{sec:comp set}
Since we assume that we always measure four Type ${\cal S}$ observables,
the magnitudes of the four transversity amplitudes, $r_i\equiv|b_i|$,
 can always be determined unambiguously. 
Three double spin observables can in general determine the relative phases
 between the four helicity amplitudes, but leave us with discrete
 ambiguities.
Therefore, more measurements are required to resolve these ambiguities.
We claim the following surprising result:
\begin{quote}
{\it In addition to the set ${\cal S}$, {\bf four} appropriately chosen
double spin observables are sufficient to determine the amplitudes uniquely.}
\end{quote}
This means that a total of {\bf eight} properly chosen measurements 
 can resolve all ambiguities.
This result contradicts the BDS rule, which asserted that {\bf nine}
 measurements
 are necessary.
In the following discussion, we first provide one explicit example
 which shows that
 eight measurements are sufficient. 
Then we present our complete results and guidelines for all situations.

Here we choose the same measurements as in the example given by
 BDS~\cite{BDS}.
Suppose that we measure $G$, $F$ and $L_x$, along with the set ${\cal S}$.
We then have the equations (see Column 3 of Table~\ref{tbl:spinob})
\begin{eqnarray}
  G  &=& - r_1 r_3 \sin(\phi_{13}) - r_2 r_4 \sin(\phi_{24}) \\
  F  &=&   r_1 r_3 \sin(\phi_{13}) - r_2 r_4 \sin(\phi_{24}) \\
 L_x &=& - r_1 r_2 \sin(\phi_{12}) - r_3 r_4 \sin(\phi_{34}) ~,
\end{eqnarray}
where we write the amplitudes $b_i=r_i\exp(\phi_i)$ and
 $\phi_{ij}=\phi_i-\phi_j$.
Except for slightly different conventions, the solutions given by
 BDS~\cite{BDS} are
\begin{eqnarray}
  \phi_{13} &=& \alpha_{13} \quad\mbox{or}\quad \pi-\alpha_{13}
  \label{eq:phi13} \\
  \phi_{24} &=& \alpha_{24} \quad\mbox{or}\quad \pi-\alpha_{24}
  \label{eq:phi24} \\
  \phi_{12} &=& \beta+\gamma \quad\mbox{or}\quad \beta+(\pi-\gamma)~,
  \label{eq:phi12} 
\end{eqnarray}
 where $\alpha_{13}$, $\alpha_{24}$, $\beta$ and $\gamma$ are defined
 by\footnote{Here $\alpha_{13}$ and $\alpha_{24}$ are uniquely defined.
    Once $\phi_{13}$ and $\phi_{24}$ (4 choices) are selected,
    $A$ is fixed and so are $\beta$ and $\gamma$.
    There are still 2 choices for $\phi_{12}$ (Eq.~(\ref{eq:phi12})).}
\begin{eqnarray}
  \sin{\alpha_{13}} &=& \frac{F-G}{2r_1 r_3}~,
    \qquad -\frac{\pi}{2} \leq \alpha_{13} \leq \frac{\pi}{2}  \nonumber \\
  \sin{\alpha_{24}} &=& - \frac{G+F}{2r_2 r_4}~,
    \qquad -\frac{\pi}{2} \leq \alpha_{24} \leq \frac{\pi}{2}  \nonumber \\
  \sin{\gamma} &=& - \frac{L_x}{A}~,
    \qquad -\frac{\pi}{2} \leq \gamma \leq \frac{\pi}{2}  \nonumber \\
  \sin{\beta} &=& \frac{r_3 r_4 \sin(\phi_{13}-\phi_{24})}{A}~,\quad
  \cos{\beta}~=~\frac{r_1r_2+r_3r_4\cos(\phi_{13}-\phi_{24})}{A}\nonumber\\
  A&=&[r_1^2r_2^2+r_3^2r_4^2+2r_1r_2r_3r_4\cos(\phi_{13}-\phi_{24})]^{1/2}~.
\end{eqnarray}
Therefore, we have an eightfold ambiguity in determining the phases.
BDS showed that two more measurements, e.g., $E$ and $L_z$, can resolve
 the ambiguity.
But, instead of two, we can show that only one additional appropriately
 chosen measurement can completely determine the four amplitudes.
Instead of the BDS choice of $E$ and $L_z,$
 take just $T_x=-r_1r_2\cos(\phi_{12})+r_3r_4\cos(\phi_{34})$ as the fourth
 double spin observable in addition to $G$, $F$ and $L_x.$ 
Using $L_x$ and $T_x$ to solve for the $\phi_{12}$ and $\phi_{34}$ phases,
 we get
\begin{equation}
  \left\{\begin{array}{l}
    \phi_{12} = - \xi + \alpha_{12} \\
    \phi_{34} = \xi + \alpha_{34}
  \end{array} \right.
  \quad \mbox{or} \quad
  \left\{\begin{array}{l}
    \phi_{12} = - \xi + (\pi-\alpha_{12}) \\
    \phi_{34} = \xi + (\pi-\alpha_{34})
  \end{array} \right. ~,
\label{eq:phi12,34}
\end{equation}
 where $\alpha_{12}$, $\alpha_{34}$ and $\xi$ are uniquely determined from
 experiment by 
\begin{eqnarray}
  \sin{\alpha_{12}} &=& - \frac{L_x^2+T_x^2+r_1^2r_2^2-r_3^2r_4^2}
      {2r_1r_2\sqrt{L_x^2+T_x^2}}~,
    \qquad -\frac{\pi}{2} \leq \alpha_{12} \leq \frac{\pi}{2}  \nonumber \\
  \sin{\alpha_{34}} &=& - \frac{L_x^2+T_x^2-r_1^2r_2^2+r_3^2r_4^2}
      {2r_1r_2\sqrt{L_x^2+T_x^2}}~,
    \qquad -\frac{\pi}{2} \leq \alpha_{34} \leq \frac{\pi}{2}  \nonumber \\
  \sin{\xi} &=& \frac{T_x}{\sqrt{L_x^2+T_x^2}}~,
   \quad \cos{\xi} \ =\ \frac{L_x}{\sqrt{L_x^2+T_x^2}}~.
\end{eqnarray}
Note that Eq.~(\ref{eq:phi12,34}) has a twofold ambiguity in determining
 $\phi_{12}$ and $\phi_{34}$, unlike the fourfold ambiguity for the
 solutions of $\phi_{13}$ and $\phi_{24}$ (Eq.~(\ref{eq:phi13}) and
 (\ref{eq:phi24})).

Combining the four solutions for $\phi_{13}$ and $\phi_{24}$
 (Eq.~(\ref{eq:phi13}) and (\ref{eq:phi24})) and the two solutions for
 $\phi_{12}$ and $\phi_{34}$ (Eq.~(\ref{eq:phi12,34})), we now have eight
 sets of possible solutions.
Using the relation $\phi_{34}=\phi_{12}+\phi_{24}-\phi_{13}$, these eight
 solutions can be expressed by:
\begin{equation}
  2\xi = \pm (\alpha_{12}-\alpha_{34})
         \pm \Biggl\{ \begin{array}{l}
                        (\alpha_{13}-\alpha_{24}) \\
                        \pi - (\alpha_{13}+\alpha_{24})
                      \end{array} ~,
\end{equation}
 here the two $\pm$ signs are independent.
Because all $\alpha$'s and $\xi$ are fixed, only one of the above eight
 solutions will hold in general, which tells us that if there is a solution,
 then it is in general a unique solution.
Therefore, in this particular case, we have shown that eight spin observables
 can resolve all ambiguities except the overall phase.
All other cases can be evaluated in the same way.
We give some guidelines in Appendix~\ref{ap:rule} and list all the situations
 (Table~\ref{tbl:num1}--\ref{tbl:num6}) for which eight measurements can
 completely determine the amplitudes.

\section{Relations from Fierz Identities}
\label{sec:fierz}
In the previous section, an elementary, albeit tedious method was used to
 determine a unique solution. 
In this section we use a totally different approach to the same problem
 of determining which set of experiments can determine the four transversity
 amplitudes without discrete ambiguities.

We know that in field theory~\cite{Fierz}, bilinear products of currents    
 obey interchange relations known as the {\it Fierz identities}.
In our problem,
 we do not deal with the four dimensional space-time, instead we have a
 four dimensional ($b_1 \cdots b_4$) amplitude space.
Nevertheless,  the properties of the Gamma matrices are characteristic of
 four dimensional space and thus hermitian versions of the Gamma matrices,
 along with all of their known properties, are of use to us. 
Their Fierz identities,  which were particularly useful in weak interactions
 studies, are obtained from the following property of the Gamma matrices:
\begin{equation}
  \Gamma^{\alpha}_{ij} \, \Gamma^{\beta}_{st} =
  C^{\alpha \beta}_{\delta \eta}\,\Gamma^{\delta}_{it}\,\Gamma^{\eta}_{sj}~,
\end{equation}
 where $C^{\alpha\beta}_{\delta\eta} \equiv \frac{1}{16}
  \mbox{Tr}(\Gamma^{\delta}\Gamma^{\alpha}\Gamma^{\eta}\Gamma^{\beta})$.
These identities are  properties of the (hermitian) Gamma matrices as
 discussed in Appendix~\ref{ap:gamma}.
We can therefore use the above {\it Fierz identities} for the Gamma
 matrices, even though we are in a context entirely different from their
 field theory origin.

Applying the Fierz transformations to the BHP forms for spin
observables,  yields the following set of relations between
observables:
\begin{eqnarray}
  \check{\Omega}^{\alpha} \, \check{\Omega}^{\beta}
  &=& (\frac{1}{2}b^{\ast}_i\widetilde{\Gamma}^{\alpha}_{ij}b_j)
      (\frac{1}{2}b^{\ast}_s\widetilde{\Gamma}^{\beta}_{st}b_t) \nonumber \\
  &=& C^{\alpha\beta}_{\delta\eta}
      (\frac{1}{2}b^{\ast}_i\widetilde{\Gamma}^{\delta}_{it}b_t)
      (\frac{1}{2}b^{\ast}_s \widetilde{\Gamma}^{\delta}_{sj}b_j) \nonumber\\
  &=& C^{\alpha\beta}_{\delta\eta} \, \check{\Omega}^{\delta}
       \, \check{\Omega}^{\eta}~,
\label{eq:fierz}
\end{eqnarray} or, since the above profile functions satisfy
$\check{\Omega}^{\alpha} \equiv \Omega^{\alpha} {\cal I},$
for all $\alpha:$
\begin{equation}
  \Omega^{\alpha} \, \Omega^{\beta} =
    C^{\alpha\beta}_{\delta\eta} \, \Omega^{\delta} \, \Omega^{\eta}~.
\label{eq:fierz2}
\end{equation}
All distinct Fierz relations derived from Eq.~(\ref{eq:fierz2}) are
 presented in Appendix~\ref{ap:fierz}.
In the rest of this section, we will show that these relations provide an
 alternate way to obtain some useful results.

\subsection{Fierz observable constraints and bounds}

The Fierz relations yield explicit and rigorous relationships
between observables.  Of course,  such relationships can be derived 
from the bilinear structure of the observables,  with much effort.
That effort is now replaced by simply invoking the
well-known Fierz rules as a general property.  That allows
us to avoid much algebra and to find all relations in one step.
There are direct  physical consequences of these relations.

For example, from Eq.~(L.tr), (L.br) and (L.bt) in Appendix~\ref{ap:fierz},
 it can be seen that if three double spin observables in a type set
 are known, then the fourth member of that type is uniquely determined.
The fourth measurement is thus redundant.

The Fierz relations can also be used to derive bounds on measurements.
For example, from Eq.~(L.tr) and (S.tr),
\begin{eqnarray}
   &(\Omega_{ 6})^2 + (\Omega_{13})^2 + (\Omega_{ 8})^2 + (\Omega_{15})^2
    \pm 2\, (\Omega_6 \,\Omega_{15} - \Omega_8 \,\Omega_{13}) \nonumber \\
  =&(\Omega_{ 1})^2 + (\Omega_{ 4})^2 - (\Omega_{10})^2 - (\Omega_{12})^2 
    \pm 2\, (\Omega_1 \Omega_{ 4} - \Omega_{10} \,\Omega_{12})
\end{eqnarray}
 we obtain
\begin{equation}
   (\Omega_{ 6} \pm \Omega_{15})^2 + (\Omega_{ 8} \mp \Omega_{13})^2
  =(\Omega_{ 1} \pm \Omega_{ 4})^2 - (\Omega_{10} \pm \Omega_{12})^2 ~.
\label{eq:bound}
\end{equation}
The left hand side of the equation is positive, so is the right hand side.
Therefore, Eq.~(\ref{eq:bound}) gives a bound relation
\begin{equation}
   \Omega_{ 1} \pm \Omega_{ 4} \geq |\Omega_{10} \pm \Omega_{12}|
   \qquad \mbox{or} \qquad
   1 \pm \Sigma \geq |T \pm P| ~.
\end{equation}
Other bounds, within the set ${\cal S}$, can be derived in the same way:
\begin{equation}
   1 \pm T \geq |P \pm \Sigma|~, \qquad 1 \pm P \geq |\Sigma \pm T| ~.
\end{equation}
Since the left hand sides of Eqs.~(S.bt), (S.br) and (S.tr) in 
 Appendix~\ref{ap:fierz} are positive,  we can deduce the bounds
\begin{eqnarray}
1 + \Sigma^2 &\geq&      P^2 + T^2  \nonumber \\
1 + T^2      &\geq& \Sigma^2 + P^2  \nonumber \\
1 + P^2      &\geq& \Sigma^2 + T^2
\end{eqnarray}
 as well as $P^2 \leq 1$, $\Sigma^2 \leq 1$ and $T^2 \leq 1$.

In this way all bounds among spin observables given by BDS~\cite{BDS} and
 Goldstein {\it et al.}~\cite{Goldstein} can be obtained using these
 Fierz relations.

\subsection{Fierz and selection of experiments}

We determined how to pick a complete set of measurements by solving
 trigonometric equations in Section~\ref{sec:comp set}.
Here we show that the Fierz relations can accomplish the same task.

Take the case of the four double spin observables $G$, $H$, $O_x$ and $C_x$
 ($\Omega_{3,\, 5,\, 14,\, 16}$).
 From Eq.~(S.bt), we can determine $(\Omega_9)^2+(\Omega_{11})^2$.
Therefore, Eq.~(S.br) and (S.b) can yield the magnitudes of $\Omega_2$
 and $\Omega_7$.
Finally, invoking Eq.~(L.br) we can uniquely decide
 $\Omega_2$ and $\Omega_7.$
By selecting the appropriate equations from Appendix~\ref{ap:fierz},
 one can determine all other observables.
Once all the observables $\check{\Omega}_{\alpha}$ are known, we can use
 $ \frac{1}{2}b^*_i \, b_j = \sum_{\alpha} \,
 \widetilde{\Gamma}_{i,j}^\alpha \ \check{\Omega}_{\alpha} $, or
$ \frac{1}{2}H^*_i \, H_j = \sum_{\alpha} \, \Gamma_{ i, j}^\alpha \ 
  \check{\Omega}_{\alpha}$, to obtain the amplitudes;  here the sum
 is over $ \alpha =1 \cdots 16$ ---  all the now known observables. 

In this specific case, we show that the chosen eight measurements
 resolve the ambiguities.
All other cases can be examined in similar way using properly
selected Fierz relations to determine the unmeasured observables,
 although sometimes it becomes rather awkward to find the right set of Fierz
 relations needed for the task.  The same result that was found earlier,
which is summarized in Tables III-VIII, are recovered by this second method.
 One advantage of the Fierz-based method is that it
provides a procedure that could possibly be generalized to reactions with
 $N >4$ amplitudes.

\section{Conclusions}
We have re-examined the classic question of how many observables
 are required in pseudoscalar meson photoproduction to completely
 and unambiguously extract the basic amplitudes from experiment. 
 The four magnitudes and three phases suggest that, aside from an overall
 arbitrary phase, only seven experiments are needed.
However, seven experiments do not suffice to resolve discrete ambiguities,
 as has been discussed, most recently by Keaton and Workman.
Stimulated by that observation, 
 we have investigated the question of the number of spin observables
 needed to determine the transversity amplitudes (assuming   
 the cross section plus all single spin observables are measured).
It is convenient to transform to transversity amplitudes, which use the
  normal to the scattering plane as the quantization axis.  
In that case,  the cross section plus the three single spin observables
 determine the magnitudes of the transversity amplitudes, while the double
 spin observables play the role of determining their phases.  
It is found that by carefully selecting four of the double spin
 observables it is possible to extract all of the requisite phases 
without discrete ambiguities.  

This is illustrated following the same procedure used in the classic BDS
 paper, by explicitly expressing observables in terms of the magnitudes and 
relative phases of the amplitudes.
As an alternate approach,  we expressed all
 observables in terms of bilinear helicity product forms, which maps 
the algebra of observables to the algebra of Hermitian  versions of
 the well known $4 \times 4$ Gamma matrices.  This mapping allows us to make
 unitary (canonical) transformations which have no effect on the observables,
 as illustrated by the transformation to the transversity basis. 
 In addition, there are transformations which can store the discrete 
ambiguities.  By finding the set of those unitary matrices that 
describe the discrete transformations, it is possible to delineate 
which experiments resolve such discrete ambiguities.  

	In addition, once it is recognized that the algebra of
 observables maps over to the algebra of the $4 \times 4$ Gamma matrices,
 we can use all known properties of the $4 \times 4$ matrices.  
In our case, the four dimensional space is not that of space-time,
 but is rather the four dimensional helicity space.  Nevertheless, 
the Gamma matrices have the well known properties of four dimensional space. 
 One property that is particularly interesting is the Fierz 
transformation.  It is shown that the  Fierz  transformation 
 properties lead to relationships between observables which can be used 
 to provide constraints and inequalities rules for  observables. 

The Fierz  transformation  can also be used as an alternate way 
to prove that a set of eight experiments can be selected to
 form a complete set of measurements.  All examples of the 4 
sets of double spin observables are presented in Table form, since
 we have not been able to express this result using a simple guideline.

\section*{Acknowledgments}
The authors wish to thank Dr. R.~Workman for his comments, 
 which stimulated us to examine this problem. 
We also wish to thank Dr. Bijan Saghai for helpful discussions.

\bigskip

\appendix

\section{Gamma Matrices}
\label{ap:gamma}
The sixteen $4\times4$ $\Gamma$ matrices are Hermitian versions of the
 familiar Dirac matrices:
\begin{equation}
  \Gamma^{\alpha=1\cdots16} = 1,\gamma^0,i\vec{\gamma},i\sigma^{0x},
  i\sigma^{0y},i\sigma^{0z},i\sigma^{xy},i\sigma^{xz},i\sigma^{zy},
  i\gamma^5\gamma^0,\gamma^5\vec{\gamma},\gamma^5.
\end{equation}
They have the following properties which are used in this paper:

(a) $\Gamma^{\alpha}$ are Hermitian and unitary.

(b) $\mbox{Tr}(\Gamma^{\alpha}\Gamma^{\beta})=4\delta_{\alpha\beta}$.

(c) $\Gamma^{\alpha}$ are linearly independent.
    Therefore, they form a complete set (a basis) for $4\times4$ matrices.
    Using Property (b), any $4\times4$ matrices $X$ can be expanded as
     $X=\sum_{\alpha}C_{\alpha}\Gamma^{\alpha}$ with $C_{\alpha}
     =\frac{1}{4}\mbox{Tr}(\Gamma^{\alpha}X)$.

(d) $\sum_{\alpha}\Gamma^{\alpha}_{ba}\Gamma^{\alpha}_{st}
      = 4\delta_{as}\delta_{bt}$.

(e) $\Gamma^{\alpha}\Gamma^{\beta}=\rho_{\alpha\beta\gamma}\Gamma^{\gamma}$
    with $\rho_{\alpha\beta\gamma}
    =\frac{1}{4}\mbox{Tr}(\Gamma^{\alpha}\Gamma^{\beta}\Gamma^{\gamma})$.

(f) $\frac{1}{4}\rho_{\alpha\gamma\delta} \rho_{\beta\gamma\eta}
    = \frac{1}{16}
      \mbox{Tr}(\Gamma^{\delta}\Gamma^{\alpha}\Gamma^{\eta}\Gamma^{\beta})
    \equiv C^{\alpha\beta}_{\delta\eta}$, which is used for the Fierz
    transformation in Section~\ref{sec:fierz}.

These properties are preserved under any unitary transformation, e.g.,
 the transversity transformation $U^{(4)}$ (defined in Eq.~(\ref{eq:u4})).
Therefore, the $\widetilde{\Gamma}$ matrices in the transversity basis have
 the same properties as the original $\Gamma$ matrices.
These sixteen $\widetilde{\Gamma}$ matrices can be grouped into four classes
 with four members in each class according to their ``shape.''
(By shape,  we mean the location of nonzero entries in $\widetilde{\Gamma}$
 matrices.)
The four shapes are: diagonal $(D)$; right parallelogram $(PR)$;
 antidiagonal $(AD)$; and left parallelogram $(PL)$~\cite{vmeson}.
In the transversity basis, these four shape classes correspond to ${\cal S}$,
 ${\cal BT}$, ${\cal BR}$ and ${\cal TR}$ type experiments.
Here, we give their explicit expressions:
\begin{equation}
\widetilde{\Gamma}_{D} = \widetilde{\Gamma}_{\cal S} =
  \left[ \begin {array}{cccc}
           a&0&0&0 \\
           0&b&0&0 \\
           0&0&c&0 \\
           0&0&0&d
         \end {array} \right] ~; \qquad
  \begin{array}{lcccc}       & a& b& c& d \\
    \widetilde{\Gamma}_{1} \ &+1&+1&+1&+1 \\
    \widetilde{\Gamma}_{4} \ &+1&+1&-1&-1 \\
    \widetilde{\Gamma}_{10}\ &-1&+1&+1&-1 \\
    \widetilde{\Gamma}_{12}\ &-1&+1&-1&+1
  \end{array}
\end{equation}

\begin{equation}
\widetilde{\Gamma}_{PR} = \widetilde{\Gamma}_{\cal BT} =
  \left[ \begin {array}{cccc}
           0&0&a&0 \\
           0&0&0&b \\
           c&0&0&0 \\
           0&d&0&0
         \end{array} \right] ~; \qquad
  \begin {array}{lcccc}      & a& b& c& d \\
    \widetilde{\Gamma}_{3} \ &-i&-i&+i&+i \\
    \widetilde{\Gamma}_{5} \ &+1&-1&+1&-1 \\
    \widetilde{\Gamma}_{9} \ &+1&+1&+1&+1 \\
    \widetilde{\Gamma}_{11}\ &+i&-i&-i&+i
  \end{array}
\end{equation}

\begin{equation}
\widetilde{\Gamma}_{AD} = \widetilde{\Gamma}_{\cal BR} =
  \left[ \begin {array}{cccc}
           0&0&0&a \\
           0&0&b&0 \\
           0&c&0&0 \\
           d&0&0&0
         \end{array} \right] ~; \qquad
  \begin {array}{ccccc}      & a& b& c& d \\
    \widetilde{\Gamma}_{14}\ &-1&+1&+1&-1 \\
    \widetilde{\Gamma}_{7} \ &-i&-i&+i&+i \\
    \widetilde{\Gamma}_{16}\ &+i&-i&+i&-i \\
    \widetilde{\Gamma}_{2} \ &+1&+1&+1&+1
  \end{array}
\end{equation}

\begin{equation}
\widetilde{\Gamma}_{PL} = \widetilde{\Gamma}_{\cal TR} =
  \left[ \begin {array}{cccc}
           0&a&0&0 \\
           b&0&0&0 \\
           0&0&0&c \\
           0&0&d&0
         \end{array} \right] ~; \qquad
  \begin{array}{lcccc}       & a& b& c& d \\
    \widetilde{\Gamma}_{6} \ &-1&-1&+1&+1 \\
    \widetilde{\Gamma}_{13}\ &+i&-i&-i&+i \\
    \widetilde{\Gamma}_{8} \ &-i&+i&-i&+i \\
    \widetilde{\Gamma}_{15}\ &-1&-1&-1&-1
  \end{array}
\end{equation}

\section{Discrete Ambiguities in Helicity Basis}
\label{ap:kw}
In Ref.~\cite{Workman}, KW gave discrete ambiguity relations associated with
 transformations of helicity amplitudes:

\centerline{Ambiguity I}
\begin{equation}
\begin{array}{lcr}
  & &  \\
  H_1 & \longleftrightarrow &  H_4 \\
  H_2 & \longleftrightarrow & -H_3 \\
  & &
\end{array}
\mbox{ ; i.e. }
\left[\begin{array}{c}
             H_1 \\
             H_2 \\
             H_3 \\
             H_4
      \end{array} \right] 
\longrightarrow
\left[\begin{array}{c}
             H'_1 \\
             H'_2 \\
             H'_3 \\
             H'_4
      \end{array} \right]
= \left[ \begin {array}{cccc}
           0& 0& 0&+1 \\
           0& 0&-1& 0 \\
           0&-1& 0& 0 \\
          +1& 0& 0& 0
         \end{array} \right]
\left[\begin{array}{c}
             H_1 \\
             H_2 \\
             H_3 \\
             H_4
      \end{array} \right]
= \Gamma^4
\left[\begin{array}{c}
             H_1 \\
             H_2 \\
             H_3 \\
             H_4
      \end{array} \right].
\end{equation}

\centerline{Ambiguity II}
\begin{equation}
\begin{array}{lcr}
  H_1 & \longrightarrow &  H_2 \\
  H_2 & \longrightarrow & -H_1 \\
  H_3 & \longrightarrow &  H_4 \\
  H_4 & \longrightarrow & -H_3
\end{array}
\mbox{ ; i.e.}
\left[\begin{array}{c}
             H_1 \\
             H_2 \\
             H_3 \\
             H_4
      \end{array} \right] 
\longrightarrow
\left[\begin{array}{c}
             H'_1 \\
             H'_2 \\
             H'_3 \\
             H'_4
      \end{array} \right]
= \left[ \begin {array}{cccc}
           0&+1& 0& 0 \\
          -1& 0& 0& 0 \\
           0& 0& 0&+1 \\
           0& 0&-1& 0
         \end{array} \right]
\left[\begin{array}{c}
             H_1 \\
             H_2 \\
             H_3 \\
             H_4
      \end{array} \right]
= -i\Gamma^{10}
\left[\begin{array}{c}
             H_1 \\
             H_2 \\
             H_3 \\
             H_4
      \end{array} \right].
\end{equation}

\centerline{Ambiguity III}
\begin{equation}
\begin{array}{lcr}
  H_1 & \longrightarrow &  H_3 \\
  H_2 & \longrightarrow &  H_4 \\
  H_3 & \longrightarrow & -H_1 \\
  H_4 & \longrightarrow & -H_2
\end{array}
\mbox{ ; i.e. }
\left[\begin{array}{c}
             H_1 \\
             H_2 \\
             H_3 \\
             H_4
      \end{array} \right] 
\longrightarrow
\left[\begin{array}{c}
             H'_1 \\
             H'_2 \\
             H'_3 \\
             H'_4
      \end{array} \right]
= \left[ \begin {array}{cccc}
           0& 0&+1& 0 \\
           0& 0& 0&+1 \\
          -1& 0& 0& 0 \\
           0&-1& 0& 0
         \end{array} \right]
\left[\begin{array}{c}
             H_1 \\
             H_2 \\
             H_3 \\
             H_4
      \end{array} \right]
= i\Gamma^{12}
\left[\begin{array}{c}
             H_1 \\
             H_2 \\
             H_3 \\
             H_4
      \end{array} \right].
\end{equation}

\centerline{Ambiguity IV}
\begin{equation}
\begin{array}{lcr}
  H_1 & \longrightarrow & -H_1^* \\
  H_2 & \longrightarrow &  H_2^* \\
  H_3 & \longrightarrow &  H_3^* \\
  H_4 & \longrightarrow & -H_4^*
\end{array}
\mbox{ ; i.e. }
\left[\begin{array}{c}
             H_1 \\
             H_2 \\
             H_3 \\
             H_4
      \end{array} \right] 
\longrightarrow
\left[\begin{array}{c}
             H'_1 \\
             H'_2 \\
             H'_3 \\
             H'_4
      \end{array} \right]
= \left[ \begin {array}{cccc}
          -1& 0& 0& 0 \\
           0&+1& 0& 0 \\
           0& 0&+1& 0 \\
           0& 0& 0&-1
         \end{array} \right]
\left[\begin{array}{c}
             H^*_1 \\
             H^*_2 \\
             H^*_3 \\
             H^*_4
      \end{array} \right]
= \Gamma^{15}
\left[\begin{array}{c}
             H^*_1 \\
             H^*_2 \\
             H^*_3 \\
             H^*_4
      \end{array} \right].
\end{equation}

Note that in the above the helicity amplitudes and Gamma matrices are in
 the original basis.
Since we work exclusively in the transversity basis in this paper, it is
 convenient to express the above ambiguities in the transversity basis:

\centerline{Ambiguity I}
\begin{equation}
\left[\begin{array}{c}
             b_1 \\
             b_2 \\
             b_3 \\
             b_4
      \end{array} \right] 
\longrightarrow
\left[\begin{array}{c}
             b'_1 \\
             b'_2 \\
             b'_3 \\
             b'_4
      \end{array} \right]
= \left[ \begin {array}{cccc}
          +1& 0& 0& 0 \\
           0&+1& 0& 0 \\
           0& 0&-1& 0 \\
           0& 0& 0&-1
         \end{array} \right]
\left[\begin{array}{c}
             b_1 \\
             b_2 \\
             b_3 \\
             b_4
      \end{array} \right]
= \widetilde{\Gamma}^4
\left[\begin{array}{c}
             b_1 \\
             b_2 \\
             b_3 \\
             b_4
      \end{array} \right]~.
\end{equation}

\centerline{Ambiguity II}
\begin{equation}
\left[\begin{array}{c}
             b_1 \\
             b_2 \\
             b_3 \\
             b_4
      \end{array} \right] 
\longrightarrow
\left[\begin{array}{c}
             b'_1 \\
             b'_2 \\
             b'_3 \\
             b'_4
      \end{array} \right]
= \left[ \begin {array}{cccc}
          -1& 0& 0& 0 \\
           0&+1& 0& 0 \\
           0& 0&+1& 0 \\
           0& 0& 0&-1
         \end{array} \right]
\left[\begin{array}{c}
             b_1 \\
             b_2 \\
             b_3 \\
             b_4
      \end{array} \right]
= -i\widetilde{\Gamma}^{10}
\left[\begin{array}{c}
             b_1 \\
             b_2 \\
             b_3 \\
             b_4
      \end{array} \right]~.
\end{equation}

\centerline{Ambiguity III}
\begin{equation}
\left[\begin{array}{c}
             b_1 \\
             b_2 \\
             b_3 \\
             b_4
      \end{array} \right] 
\longrightarrow
\left[\begin{array}{c}
             b'_1 \\
             b'_2 \\
             b'_3 \\
             b'_4
      \end{array} \right]
= \left[ \begin {array}{cccc}
          -1& 0& 0& 0 \\
           0&+1& 0& 0 \\
           0& 0&-1& 0 \\
           0& 0& 0&+1
         \end{array} \right]
\left[\begin{array}{c}
             b_1 \\
             b_2 \\
             b_3 \\
             b_4
      \end{array} \right]
= i\widetilde{\Gamma}^{12}
\left[\begin{array}{c}
             b_1 \\
             b_2 \\
             b_3 \\
             b_4
      \end{array} \right]~.
\end{equation}

\centerline{Ambiguity IV}
\begin{equation}
\left[\begin{array}{c}
             b_1 \\
             b_2 \\
             b_3 \\
             b_4
      \end{array} \right] 
\longrightarrow
\left[\begin{array}{c}
             b'_1 \\
             b'_2 \\
             b'_3 \\
             b'_4
      \end{array} \right]
= \left[ \begin {array}{cccc}
           0&-1& 0& 0 \\
          -1& 0& 0& 0 \\
           0& 0& 0&-1 \\
           0& 0&-1& 0
         \end{array} \right]
\left[\begin{array}{c}
             b^*_1 \\
             b^*_2 \\
             b^*_3 \\
             b^*_4
      \end{array} \right]
= \widetilde{\Gamma}^{15}
\left[\begin{array}{c}
             b^*_1 \\
             b^*_2 \\
             b^*_3 \\
             b^*_4
      \end{array} \right]~.
\end{equation}

Ambiguity~I, II and III are equivalent to our linear ambiguity
 $L=\widetilde{\Gamma}^4$, $\widetilde{\Gamma}^{10}$, and
 $\widetilde{\Gamma}^{12}$ except for irrelevant phases
 (see Table~\ref{tbl:amb}).
Ambiguity~IV corresponds to our antilinear ambiguity
 $A=\widetilde{\Gamma}^{15}$.
And the other three antilinear ambiguities in Table~\ref{tbl:amb},
 $A=\widetilde{\Gamma}^6$, $\widetilde{\Gamma}^{13}$, and
 $\widetilde{\Gamma}^8$, can be constructed by Ambiguity IV and the three
 linear ambiguity (Ambiguity I to III).
It is shown explicitly by
\begin{eqnarray*}
  \widetilde{\Gamma}^6 &=& \widetilde{\Gamma}^4 \widetilde{\Gamma}^{15} \\
  \widetilde{\Gamma}^{13}&=&i\widetilde{\Gamma}^{10}\widetilde{\Gamma}^{15}\\
  \widetilde{\Gamma}^8 &=& -i\widetilde{\Gamma}^{12}\widetilde{\Gamma}^{15}~.
\end{eqnarray*}
Here we recover the results given by KW (Ref.~\cite{Workman}).

\section{Complete Sets of Eight Measurements}
\label{ap:rule}
Here we give rules for choosing four double spin observables which can
 resolve the ambiguities when they are taken together with $\sigma(\theta)$,
 $\Sigma$, $T$ and $P$.
These rules are not expressed succinctly, and we can not yet provide
simple physical guidance.
Some may find Table~III--VIII also useful for choosing the
 appropriate measurements.

Define ${\cal A,\ B,\ C,\ D,\ E}$ and $\cal F$ as sets of pairs of double
 spin observables:
\begin{eqnarray}
  \{(H,E),\,(O_x,C_z),\,(T_x,L_z)\} &=& {\cal A} \\
  \{(G,F),\,(O_z,C_x),\,(T_z,L_x)\} &=& {\cal B} \\
  \{(H,F),\,(O_x,C_x),\,(T_x,T_z)\} &=& {\cal C} \\
  \{(G,H),\,(O_x,O_z),\,(T_x,L_x)\} &=& {\cal D} \\
  \{(E,F),\,(C_x,C_z),\,(T_z,L_z)\} &=& {\cal E} \\
  \{(G,E),\,(O_z,C_z),\,(L_x,L_z)\} &=& {\cal F} ~,
\end{eqnarray}
 and $\cal X$ and $\cal Y$ as sets of double spin observables:
\begin{eqnarray}
  \{H,\,E,\,O_x,\,C_z,\,T_x,\,L_z\} &=& {\cal X} \\
  \{G,\,F,\,O_z,\,C_x,\,T_z,\,L_x\} &=& {\cal Y} ~.
\end{eqnarray}

There are four situations in choosing four double spin observables: \\
(a) \underline{2 + 2} cases: Pick one pair of double spin observables from
     the same type ($\cal BT$, $\cal BR$ or $\cal TR$), and another pair from
     another type.
    Here are the 2 + 2 cases which can determine the amplitudes uniquely:
\begin{enumerate}
\item  2 $\cal BT$ + 2 $\cal TR$ cases:
       At least one pair belongs to set $\cal D$ or $\cal E$, i.e., at least
        one pair is $(G,H)$, $(E,F)$, $(T_x,L_x)$ or $(T_z,L_z)$.
\item  2 $\cal BT$ + 2 $\cal BR$ cases:
       At least one pair belongs to set $\cal C$ or $\cal F$, i.e., at least
        one pair is $(G,E)$, $(H,F)$, $(O_x,C_x)$ or $(O_z,C_z)$.
\item  2 $\cal BR$ + 2 $\cal TR$ cases:
       The $\cal BR$ pair belongs to set $\cal D$ or $\cal E$, or the
        $\cal TR$ pair belongs to set ${\cal C}$ or ${\cal F}$, i.e.,
        at least one pair is $(O_x,O_z)$, $(C_x,C_z)$, $(T_x,T_z)$ or
        $(L_x,L_z)$.
\end{enumerate}
(b) \underline{2 + 1 + 1} cases: Pick one pair of double spin observables
     from one type of $\cal BT$, $\cal BR$ or $\cal TR$, and one observable
     from each of the remaining two types.
    The followings are the only 2 + 1 + 1 situations under which the
     ambiguities are \underline{not} resolved:
\begin{enumerate}
\item  When the pair belongs to set $\cal A$ and the other two observables
       belong to the same set of $\cal X$ or $\cal Y$.
\item  When the pair belongs to set $\cal B$ and the other two observables
       belong to the different set of $\cal X$ and $\cal Y$.
\end{enumerate}
(c) \underline{3 + 1} cases: Pick 3 double spin observables from one type
     and one observable from other types.
    Ambiguities can not be resolved in these cases. \\
(d) \underline{4} cases: Pick all 4 double spin observables from the same
     type.
    They can never determine the amplitudes uniquely.

\section{Fierz Relations}
\label{ap:fierz}
 In this appendix, we display all of the Fierz relations
as contained in Eq.~(\ref{eq:fierz2}).  We select values
for the index $\alpha\ \&\  \beta$ and then evaluate
the coefficient
$C^{\alpha \beta}_{\delta \eta},$ from the
trace rules in Appendix~\ref{ap:gamma}.  There are
$16 \times 16 =256$ choices for the pair $\alpha\ , \beta$;
however, due to symmetries and the fact that many of the resulting
equations are redundant,  we can reduce the Fierz results
to the following 37 equations.  There continue to be some
redundancy in these,  but that is hard to judge since they are
nonlinear equations.

 The 37  surviving equations that are obtained using the
Fierz identities are organized 
according to the following scheme: 

a. Relations (L.tr) and (S.tr) involve only ${\cal S}$ and ${\cal TR}$
   types, etc.;

b. Relations (Q.b) and (S.b) involve only ${\cal BT}$ and ${\cal BR}$
   types, etc.; 

c. Relations (Q.bt.1-4), (Q.br.1-4), (Q.tr.1-4) and (L.1--12) involve all
   four $\cal TR$, $\cal BT$, $\cal BR$, and $\cal S$ types.

The label ``L'' is used for Linear (LHS) to Quadratic (RHS) relations
 (with $\alpha=1$ selected).
The label ``Q" denotes purely quadratic relations on right and left sides,
 while ``S" is used to indicate observable squared rules.
Use Table~\ref{tbl:spinob} to translate these
rules to the notation $(\sigma, \Sigma, T, P)\,{\cal S};\ ( G, H, E, F)\,
 {\cal BT};\ ( O_x, O_z, C_x, C_z)\,{\cal BR};\ (T_x, T_z, L_x, L_z)\,
 {\cal TR}.$
Note that these relations hold true at all energies and all angles.
The type ${\cal S}$ observables $\Omega_{1, 4,10,12}$ are all assumed to
 be known.

The following relations have been organized to help in seeking connections
 between observables,  which is important in showing that not all
 observables need to be measured.
They have interesting patterns which might help in examining questions of
 interdependence between observables and in future
 generalizations to reactions with $N>4$ amplitudes.

Linear-Quadratic Relations:
\begin{eqnarray}
\Omega_{1\ } &=& \ 1 \ = \frac{1}{4}\sum_{\alpha=1}^{16} (\Omega_{\alpha})^2
  \eqnum{L.0} \\
\Omega_{4\ } &=&  \Omega_{10 } \Omega_{12}  + \Omega_6 \Omega_{15}
                - \Omega_8 \Omega_{13}  \eqnum{L.tr} \\
\Omega_{10 } &=&  \Omega_{4\ } \Omega_{12} + \Omega_2 \Omega_{14} 
                + \Omega_7 \Omega_{16}  \eqnum{L.br} \\
\Omega_{12 } &=&  \Omega_{4\ } \Omega_{10} + \Omega_3 \Omega_{11}
                - \Omega_5 \Omega_{9 }  \eqnum{L.bt} \\
\nonumber \\
\Omega_{3\ } &=& + \Omega_{11 } \Omega_{12} - \Omega_7 \Omega_{15}
                 + \Omega_{14 } \Omega_{8\ }  \eqnum{L.1} \\
\Omega_{5\ } &=& - \Omega_{9\ } \Omega_{12} + \Omega_7 \Omega_{13}
                 - \Omega_{14 } \Omega_{6\ }  \eqnum{L.2} \\
\Omega_{9\ } &=& - \Omega_{5\ } \Omega_{12} - \Omega_2 \Omega_{15} 
                 - \Omega_{16 } \Omega_{8\ }  \eqnum{L.3} \\
\Omega_{11 } &=& + \Omega_{3\ } \Omega_{12} + \Omega_2 \Omega_{13} 
                 + \Omega_{16 } \Omega_{6\ }  \eqnum{L.4} \\
\nonumber \\
\Omega_{14 } &=&  \Omega_{2\ } \Omega_{10} + \Omega_3 \Omega_{8\ }
                - \Omega_{5\ } \Omega_{6\ }  \eqnum{L.5} \\
\Omega_{7\ } &=&  \Omega_{16 } \Omega_{10} - \Omega_3 \Omega_{15 }
                + \Omega_{5\ } \Omega_{13 }  \eqnum{L.6} \\
\Omega_{16 } &=&  \Omega_{7\ } \Omega_{10} - \Omega_9 \Omega_{8\ } 
                + \Omega_{11 } \Omega_{6\ }  \eqnum{L.7} \\
\Omega_{2\ } &=&  \Omega_{14 } \Omega_{10} - \Omega_9 \Omega_{15 } 
                + \Omega_{11 } \Omega_{13 }  \eqnum{L.8} \\
\nonumber \\
\Omega_{6\ } &=& + \Omega_{15 } \Omega_4 - \Omega_{5\ } \Omega_{14 }
                 + \Omega_{11 } \Omega_{16 } \eqnum{L.9} \\
\Omega_{13 } &=& - \Omega_{8\ } \Omega_4 + \Omega_{5\ } \Omega_{7\ }
                 + \Omega_{11 } \Omega_{2\ } \eqnum{L.10} \\
\Omega_{8\ } &=& - \Omega_{13 } \Omega_4 + \Omega_{3\ } \Omega_{14 }
                 - \Omega_{9\ } \Omega_{16 } \eqnum{L.11} \\
\Omega_{15 } &=& + \Omega_{6\ } \Omega_4 - \Omega_{3\ } \Omega_{7\ }
                 - \Omega_{9\ } \Omega_{2\ } \eqnum{L.12}
\end{eqnarray}

Quadratic Relations:
\begin{eqnarray}
\Omega_{2\ } \Omega_{7\ } - \Omega_{14 } \Omega_{16 }
  - \Omega_{3\ } \Omega_{9\ } - \Omega_{5\ } \Omega_{11 }
  &=& 0 \eqnum{Q.b} \\
\Omega_{3\ } \Omega_{5\ } + \Omega_{9\ } \Omega_{11 }
  + \Omega_{6\ } \Omega_{8\ } + \Omega_{13 } \Omega_{15 }
 &=& 0 \eqnum{Q.t} \\
\Omega_{2\ } \Omega_{16 } - \Omega_{7\ } \Omega_{14 }
  - \Omega_{6\ } \Omega_{13 } - \Omega_{8\ } \Omega_{15 }
 &=& 0 \eqnum{Q.r} \\
\nonumber \\
\Omega_{4\ } \Omega_{3\ } - \Omega_{10 } \Omega_{11 }
  + \Omega_{7\ } \Omega_{6\ } + \Omega_{14 } \Omega_{13 }
  &=& 0 \eqnum{Q.bt.1} \\
\Omega_{4\ } \Omega_{5\ } + \Omega_{10 } \Omega_{9\ }
  + \Omega_{7\ } \Omega_{8\ } + \Omega_{14 } \Omega_{15 }
  &=& 0 \eqnum{Q.bt.2} \\
\Omega_{4\ } \Omega_{9\ } + \Omega_{10 } \Omega_{5\ }
  + \Omega_{2\ } \Omega_{6\ } - \Omega_{16 } \Omega_{13 }
  &=& 0 \eqnum{Q.bt.3} \\
\Omega_{4\ } \Omega_{11 } - \Omega_{10 } \Omega_{3\ }
  + \Omega_{2\ } \Omega_{8\ } - \Omega_{16 } \Omega_{15 }
  &=& 0 \eqnum{Q.bt.4} \\
\nonumber \\
\Omega_{4\ } \Omega_{14 } - \Omega_{12 } \Omega_{2\ }
  + \Omega_{3\ } \Omega_{13 } + \Omega_{5\ } \Omega_{15 }
  &=& 0 \eqnum{Q.br.1} \\
\Omega_{4\ } \Omega_{7\ } - \Omega_{12 } \Omega_{16 }
  + \Omega_{3\ } \Omega_{6\ } + \Omega_{5\ } \Omega_{8\ }
  &=& 0 \eqnum{Q.br.2} \\
\Omega_{4\ } \Omega_{16 } - \Omega_{12 } \Omega_{7\ }
  - \Omega_{9\ } \Omega_{13 } - \Omega_{11 } \Omega_{15 }
  &=& 0 \eqnum{Q.br.3} \\
\Omega_{4\ } \Omega_{2\ } - \Omega_{12 } \Omega_{14 }
  + \Omega_{9\ } \Omega_{6\ } + \Omega_{11 } \Omega_{8\ }
  &=& 0 \eqnum{Q.br.4} \\
\nonumber \\
\Omega_{10 } \Omega_{6\ } - \Omega_{12 } \Omega_{15 }
  + \Omega_{5\ } \Omega_{2\ } - \Omega_{11 } \Omega_{7\ }
  &=& 0 \eqnum{Q.tr.1} \\
\Omega_{10 } \Omega_{13 } + \Omega_{12 } \Omega_{8\ }
  - \Omega_{5\ } \Omega_{16 } - \Omega_{11 } \Omega_{14 }
  &=& 0 \eqnum{Q.tr.2} \\
\Omega_{10 } \Omega_{8\ } + \Omega_{12 } \Omega_{13 }
  - \Omega_{3\ } \Omega_{2\ } + \Omega_{9\ } \Omega_{7\ }
  &=& 0 \eqnum{Q.tr.3} \\
\Omega_{10 } \Omega_{15 } - \Omega_{12 } \Omega_{6\ }
  + \Omega_{3\ } \Omega_{16 } + \Omega_{9\ } \Omega_{14 }
  &=& 0 \eqnum{Q.tr.4}
\end{eqnarray}

Square Relations:
\begin{eqnarray}
(\Omega_{3\ })^2 + (\Omega_{5\ })^2 + (\Omega_{9\ })^2 + (\Omega_{11 })^2
&=& (\Omega_{1\ })^2 - (\Omega_{4\ })^2 - (\Omega_{10 })^2 + (\Omega_{12 })^2
  \eqnum{S.bt} \\
(\Omega_{14 })^2 + (\Omega_{7\ })^2 + (\Omega_{16 })^2 + (\Omega_{2\ })^2
&=& (\Omega_{1\ })^2 - (\Omega_{4\ })^2 + (\Omega_{10 })^2 - (\Omega_{12 })^2
  \eqnum{S.br} \\
(\Omega_{6\ })^2 + (\Omega_{13 })^2 + (\Omega_{8\ })^2 + (\Omega_{15 })^2
&=& (\Omega_{1\ })^2 + (\Omega_{4\ })^2 - (\Omega_{10 })^2 - (\Omega_{12 })^2
  \eqnum{S.tr} \\
\nonumber \\
(\Omega_{3\ })^2 + (\Omega_{5\ })^2 - (\Omega_{9\ })^2 - (\Omega_{11 })^2
&=& (\Omega_{14 })^2 + (\Omega_{7\ })^2 - (\Omega_{16 })^2 - (\Omega_{2\ })^2
  \eqnum{S.b} \\
-(\Omega_{3\ })^2 + (\Omega_{5\ })^2 - (\Omega_{9\ })^2 + (\Omega_{11 })^2
&=& (\Omega_{6\ })^2 + (\Omega_{13 })^2 - (\Omega_{8\ })^2 - (\Omega_{15 })^2
  \eqnum{S.t} \\
(\Omega_{14 })^2 - (\Omega_{7\ })^2 + (\Omega_{16 })^2 - (\Omega_{2\ })^2
&=& (\Omega_{6\ })^2 - (\Omega_{13 })^2 + (\Omega_{8\ })^2 - (\Omega_{15 })^2
  \eqnum{S.r} 
\end{eqnarray}

\begin{table}
\squeezetable
\caption{{\bf Spin Observables:} The 16 spin observables are expressed in
 helicity representation and BHP forms.
 Here they are classified into four type sets: type ${\cal S}$ for the
 differential cross section and single spin observables, and types
 ${\cal BT}$, ${\cal BR}$ and ${\cal TR}$ for beam-target, beam-recoil and
 target-recoil spin observables,
 respectively.}
\label{tbl:spinob}
\vspace{4mm}
\begin{tabular}{l@{$\equiv$}ccccc}
  \multicolumn{2}{c}{Spin}   &    Helicity    &  Transversity  & BHP & Set\\
  \multicolumn{2}{c}{Observable} & Representation & Representation & Form
  & \hspace*{0.5in}  \\
\tableline
\multicolumn{2}{c}{} &  &  &  &  \\
$\check{\Omega}^1$     & ${\cal I}(\theta)$
  & $\frac{1}{2}(|H_1|^2 + |H_2|^2 + |H_3|^2 + |H_4|^2)$
  & $\frac{1}{2}(|b_1|^2 + |b_2|^2 + |b_3|^2 + |b_4|^2)$
  & $\frac{1}{2}\langle b|\widetilde{\Gamma}^{1}|b\rangle$ &  \\
$\check{\Omega}^4$     & $\check{\Sigma}$  & Re$(- H_1 H_4^* + H_2 H_3^*)$
  & $\frac{1}{2}(|b_1|^2 + |b_2|^2 - |b_3|^2 - |b_4|^2)$
  & $\frac{1}{2}\langle b|\widetilde{\Gamma}^{4}|b\rangle$ & ${\cal S}$  \\
$\check{\Omega}^{10}$  & $-\check{T}$    & Im$(H_1 H_2^* + H_3 H_4^*)$
  & $\frac{1}{2}(-|b_1|^2 + |b_2|^2 + |b_3|^2 - |b_4|^2)$
  & $\frac{1}{2}\langle b|\widetilde{\Gamma}^{10}|b\rangle$ &  \\
$\check{\Omega}^{12}$  & $\check{P}$     & Im$(- H_1 H_3^* - H_2 H_4^*)$
  & $\frac{1}{2}(-|b_1|^2 + |b_2|^2 - |b_3|^2 + |b_4|^2)$
  & $\frac{1}{2}\langle b|\widetilde{\Gamma}^{12}|b\rangle$ &  \\
\multicolumn{2}{c}{} &  &  &  &  \\
$\check{\Omega}^3$     & $\check{G}$     & Im$(H_1 H_4^* - H_3 H_2^*)$
  & Im$(- b_1 b_3^* - b_2 b_4^*)$
  & $\frac{1}{2}\langle b|\widetilde{\Gamma}^{3}|b\rangle$ &  \\
$\check{\Omega}^5$     & $\check{H}$     & Im$(- H_2 H_4^* + H_1 H_3^*)$
  & Re$(b_1 b_3^* - b_2 b_4^*)$
  & $\frac{1}{2}\langle b|\widetilde{\Gamma}^{5}|b\rangle$ & ${\cal BT}$  \\
$\check{\Omega}^9$     & $\check{E}$
  & $\frac{1}{2}(|H_1|^2 - |H_2|^2 + |H_3|^2 - |H_4|^2)$
  & Re$(b_1 b_3^* + b_2 b_4^*)$
  & $\frac{1}{2}\langle b|\widetilde{\Gamma}^{9}|b\rangle$ &  \\
$\check{\Omega}^{11}$  & $\check{F}$     & Re$(- H_2 H_1^* - H_4 H_3^*)$
  & Im$(b_1 b_3^* - b_2 b_4^*)$
  & $\frac{1}{2}\langle b|\widetilde{\Gamma}^{11}|b\rangle$ &  \\
\multicolumn{2}{c}{} &  &  &  &  \\
$\check{\Omega}^{14}$  & $\check{O_x}$   & Im$(- H_2 H_1^* + H_4 H_3^*)$
  & Re$(- b_1 b_4^* + b_2 b_3^*)$
  & $\frac{1}{2}\langle b|\widetilde{\Gamma}^{14}|b\rangle$ &  \\
$\check{\Omega}^7$     & $-\check{O_z}$  & Im$(H_1 H_4^* - H_2 H_3^*)$
  & Im$(- b_1 b_4^* - b_2 b_3^*)$
  & $\frac{1}{2}\langle b|\widetilde{\Gamma}^{7}|b\rangle$ & ${\cal BR}$  \\
$\check{\Omega}^{16}$  & $-\check{C_x}$  & Re$(H_2 H_4^* + H_1 H_3^*)$
  & Im$(b_1 b_4^* - b_2 b_3^*)$
  & $\frac{1}{2}\langle b|\widetilde{\Gamma}^{16}|b\rangle$ &  \\
$\check{\Omega}^2$     & $-\check{C_z}$
  & $\frac{1}{2}(|H_1|^2 + |H_2|^2 - |H_3|^2 - |H_4|^2)$
  & Re$(b_1 b_4^* + b_2 b_3^*)$
  & $\frac{1}{2}\langle b|\widetilde{\Gamma}^{2}|b\rangle$ &  \\
\multicolumn{2}{c}{} &  &  &  &  \\
$\check{\Omega}^6$     & $-\check{T_x}$  & Re$(- H_1 H_4^* - H_2 H_3^*)$
  & Re$(- b_1 b_2^* + b_3 b_4^*)$
  & $\frac{1}{2}\langle b|\widetilde{\Gamma}^{6}|b\rangle$ &  \\
$\check{\Omega}^{13}$  & $-\check{T_z}$  & Re$(- H_1 H_2^* + H_4 H_3^*)$
  & Im$(b_1 b_2^* - b_3 b_4^*)$
  & $\frac{1}{2}\langle b|\widetilde{\Gamma}^{13}|b\rangle$ & ${\cal TR}$  \\
$\check{\Omega}^8$     & $\check{L_x}$   & Re$(H_2 H_4^* - H_1 H_3^*)$
  & Im$(- b_1 b_2^* - b_3 b_4^*)$
  & $\frac{1}{2}\langle b|\widetilde{\Gamma}^{8}|b\rangle$ &  \\
$\check{\Omega}^{15}$  & $\check{L_z}$
  & $\frac{1}{2}(-|H_1|^2 + |H_2|^2 + |H_3|^2 - |H_4|^2)$
  & Re$(- b_1 b_2^* - b_3 b_4^*)$
  & $\frac{1}{2}\langle b|\widetilde{\Gamma}^{15}|b\rangle$ &  \\
\multicolumn{2}{c}{} &  &  &  &  \\
\end{tabular}
\end{table}

\begin{table}
\caption{Result of linear ($L$) and antilinear ($A$) ambiguity
 transformations applied to observables.
The observables are either invariant (+) or change sign ($-$)under
 these transformations.}
\label{tbl:amb}
\vspace*{4mm}
\begin{tabular}{ccccccccc}
Spin & \multicolumn{3}{c}{Linear Transformation $L$} &
  \multicolumn{4}{c}{Antilinear Transformation $A$} &  \\
 & \multicolumn{3}{c}{$b_i \longrightarrow b'_i = L_{ij} b_j$}
 & \multicolumn{4}{c}{$b_i \longrightarrow b'_i = A_{ij} b^{\ast}_j$}& Set\\
\cline{2-4}
\cline{5-8}
Observable & $\widetilde{\Gamma}_{4}$ & $\widetilde{\Gamma}_{10}$ &
 $\widetilde{\Gamma}_{12}$ & $\widetilde{\Gamma}_6$ & $\widetilde{\Gamma}_8$
 & $\widetilde{\Gamma}_{13}$ & $\widetilde{\Gamma}_{15}$ &  \\
\hline
 & \hspace*{0.6in} & \hspace*{0.6in} & \hspace*{0.6in} & \hspace*{0.6in}
 & \hspace*{0.6in} & \hspace*{0.6in} & \hspace*{0.6in} & \hspace*{0.6in} \\
$\sigma (\theta)$ & + & + & + & + & + & + & + &  \\
$\Sigma$          & + & + & + & + & + & + & + & ${\cal S}$ \\
$T$               & + & + & + & + & + & + & + &  \\
$P$               & + & + & + & + & + & + & + &  \\
 & & & & & & & & \\
$G$               &$-$&$-$& + &$+$&$-$&$+$&$-$&  \\
$H$               &$-$&$-$& + &$-$&$+$&$-$&$+$& ${\cal BT}$ \\
$E$               &$-$&$-$& + &$-$&$+$&$-$&$+$&  \\
$F$               &$-$&$-$& + &$+$&$-$&$+$&$-$&  \\
 & & & & & & & & \\
$O_x$             &$-$& + &$-$&$-$&$-$&$+$&$+$&  \\
$O_z$             &$-$& + &$-$&$+$&$+$&$-$&$-$& ${\cal BR}$ \\
$C_x$             &$-$& + &$-$&$+$&$+$&$-$&$-$&  \\
$C_z$             &$-$& + &$-$&$-$&$-$&$+$&$+$&  \\
 & & & & & & & & \\
$T_x$             & + &$-$&$-$&$+$&$-$&$-$&$+$&  \\
$T_z$             & + &$-$&$-$&$-$&$+$&$+$&$-$& ${\cal TR}$ \\
$L_x$             & + &$-$&$-$&$-$&$+$&$+$&$-$&  \\
$L_z$             & + &$-$&$-$&$+$&$-$&$-$&$+$&  \\
 & & & & & & & & \\
\end{tabular}
\end{table}

\begin{table}
\squeezetable
\caption{ Tables III--VIII enumerate all situations under
  which four double spin observables, along with the set ${\cal S}$, 
can completely
  determine the transversity amplitudes.
 In these tables, `{\bf X}'s' indicate three initially selected measurements,
  and `{\bf O}'s' indicate the possible choices
  for fourth observable that can resolve all the ambiguities.}
\label{tbl:num1}
\vspace*{4mm}
\begin{tabular}{c*{28}{c}c}
 & & & & & & & & & & & & & & & & & & & & & & & & & & & & &  \\
$G$  & &{\bf X}&{\bf X}&{\bf X}&{\bf X}&{\bf X}&{\bf X}&{\bf X}&{\bf X}&
       &{\bf X}&{\bf X}&{\bf X}&{\bf X}&{\bf X}&{\bf X}&{\bf X}&{\bf X}&
       &{\bf X}&{\bf X}&{\bf X}&{\bf X}&{\bf X}&{\bf X}&{\bf X}&{\bf X}& & \\
$H$  & &{\bf X}&{\bf X}&{\bf X}&{\bf X}&{\bf X}&{\bf X}&{\bf X}&{\bf X}&
       & & & & & & & & & & & & & & & & & & &${\cal BT}$ \\
$E$  & & & & & & & & & &
       &{\bf X}&{\bf X}&{\bf X}&{\bf X}&{\bf X}&{\bf X}&{\bf X}&{\bf X}&
       & & & & & & & & & &  \\
$F$  & & & & & & & & & & & & & & & & & & &
       &{\bf X}&{\bf X}&{\bf X}&{\bf X}&{\bf X}&{\bf X}&{\bf X}&{\bf X}& & \\
 & & & & & & & & & & & & & & & & & & & & & & & & & & & & &  \\
$O_x$& &{\bf X}& &{\bf O}& &{\bf O}&{\bf O}&{\bf O}&{\bf O}&
       &{\bf X}&{\bf O}&{\bf O}&{\bf O}&{\bf O}&{\bf O}&{\bf O}&{\bf O}&
       &{\bf X}& &{\bf O}& &{\bf O}& & &{\bf O}& &  \\
$O_z$& & &{\bf X}& &{\bf O}&{\bf O}&{\bf O}&{\bf O}&{\bf O}&
       &{\bf O}&{\bf X}&{\bf O}&{\bf O}&{\bf O}&{\bf O}&{\bf O}&{\bf O}&
       & &{\bf X}& &{\bf O}& &{\bf O}&{\bf O}& & &${\cal BR}$ \\
$C_x$& &{\bf O}& &{\bf X}& &{\bf O}&{\bf O}&{\bf O}&{\bf O}&
       &{\bf O}&{\bf O}&{\bf X}&{\bf O}&{\bf O}&{\bf O}&{\bf O}&{\bf O}&
       &{\bf O}& &{\bf X}& & &{\bf O}&{\bf O}& & &  \\
$C_z$& & &{\bf O}& &{\bf X}&{\bf O}&{\bf O}&{\bf O}&{\bf O}&
       &{\bf O}&{\bf O}&{\bf O}&{\bf X}&{\bf O}&{\bf O}&{\bf O}&{\bf O}&
       & &{\bf O}& &{\bf X}&{\bf O}& & &{\bf O}& &  \\
 & & & & & & & & & & & & & & & & & & & & & & & & & & & & &  \\
$T_x$& &{\bf O}&{\bf O}&{\bf O}&{\bf O}&{\bf X}&{\bf O}&{\bf O}&{\bf O}&
       &{\bf O}&{\bf O}&{\bf O}&{\bf O}&{\bf X}& &{\bf O}& &
       &{\bf O}& & &{\bf O}&{\bf X}& &{\bf O}& & &  \\
$T_z$& &{\bf O}&{\bf O}&{\bf O}&{\bf O}&{\bf O}&{\bf X}&{\bf O}&{\bf O}&
       &{\bf O}&{\bf O}&{\bf O}&{\bf O}& &{\bf X}& &{\bf O}&
       & &{\bf O}&{\bf O}& & &{\bf X}& &{\bf O}& &${\cal TR}$ \\
$L_x$& &{\bf O}&{\bf O}&{\bf O}&{\bf O}&{\bf O}&{\bf O}&{\bf X}&{\bf O}&
       &{\bf O}&{\bf O}&{\bf O}&{\bf O}&{\bf O}& &{\bf X}& &
       & &{\bf O}&{\bf O}& &{\bf O}& &{\bf X}& & &  \\
$L_z$& &{\bf O}&{\bf O}&{\bf O}&{\bf O}&{\bf O}&{\bf O}&{\bf O}&{\bf X}&
       &{\bf O}&{\bf O}&{\bf O}&{\bf O}& &{\bf O}& &{\bf X}&
       &{\bf O}& & &{\bf O}& &{\bf O}& &{\bf X}& &  \\
 & & & & & & & & & & & & & & & & & & & & & & & & & & & & &
\end{tabular}
\end{table}
\begin{table}
\squeezetable
\caption{}
\label{tbl:num2}
\vspace*{4mm}
\begin{tabular}{c*{28}{c}c}
 & & & & & & & & & & & & & & & & & & & & & & & & & & & & &  \\
$G$  & & & & & & & & & & & & & & & & & & & & & & & & & & & & &  \\
$H$  & &{\bf X}&{\bf X}&{\bf X}&{\bf X}&{\bf X}&{\bf X}&{\bf X}&{\bf X}&
       &{\bf X}&{\bf X}&{\bf X}&{\bf X}&{\bf X}&{\bf X}&{\bf X}&{\bf X}&
 & & & & & & & & & &${\cal BT}$\\
$E$  & &{\bf X}&{\bf X}&{\bf X}&{\bf X}&{\bf X}&{\bf X}&{\bf X}&{\bf X}&
       & & & & & & & & &
       &{\bf X}&{\bf X}&{\bf X}&{\bf X}&{\bf X}&{\bf X}&{\bf X}&{\bf X}& & \\
$F$  & & & & & & & & & &
       &{\bf X}&{\bf X}&{\bf X}&{\bf X}&{\bf X}&{\bf X}&{\bf X}&{\bf X}&
       &{\bf X}&{\bf X}&{\bf X}&{\bf X}&{\bf X}&{\bf X}&{\bf X}&{\bf X}& & \\
 & & & & & & & & & & & & & & & & & & & & & & & & & & & & &  \\
$O_x$& &{\bf X}& &{\bf O}& & &{\bf O}&{\bf O}& &
       &{\bf X}&{\bf O}&{\bf O}&{\bf O}&{\bf O}&{\bf O}&{\bf O}&{\bf O}&
       &{\bf X}& &{\bf O}& &{\bf O}&{\bf O}&{\bf O}&{\bf O}& &  \\
$O_z$& & &{\bf X}& &{\bf O}&{\bf O}& & &{\bf O}&
       &{\bf O}&{\bf X}&{\bf O}&{\bf O}&{\bf O}&{\bf O}&{\bf O}&{\bf O}&
       & &{\bf X}& &{\bf O}&{\bf O}&{\bf O}&{\bf O}&{\bf O}& &${\cal BR}$\\
$C_x$& &{\bf O}& &{\bf X}& &{\bf O}& & &{\bf O}&
       &{\bf O}&{\bf O}&{\bf X}&{\bf O}&{\bf O}&{\bf O}&{\bf O}&{\bf O}&
       &{\bf O}& &{\bf X}& &{\bf O}&{\bf O}&{\bf O}&{\bf O}& &  \\
$C_z$& & &{\bf O}& &{\bf X}& &{\bf O}&{\bf O}& &
       &{\bf O}&{\bf O}&{\bf O}&{\bf X}&{\bf O}&{\bf O}&{\bf O}&{\bf O}&
       & &{\bf O}& &{\bf X}&{\bf O}&{\bf O}&{\bf O}&{\bf O}& &  \\
 & & & & & & & & & & & & & & & & & & & & & & & & & & & & &  \\
$T_x$& & &{\bf O}&{\bf O}& &{\bf X}& &{\bf O}& &
       &{\bf O}&{\bf O}&{\bf O}&{\bf O}&{\bf X}& &{\bf O}& &
       &{\bf O}&{\bf O}&{\bf O}&{\bf O}&{\bf X}&{\bf O}&{\bf O}&{\bf O}& & \\
$T_z$& &{\bf O}& & &{\bf O}& &{\bf X}& &{\bf O}&
       &{\bf O}&{\bf O}&{\bf O}&{\bf O}& &{\bf X}& &{\bf O}&
       &{\bf O}&{\bf O}&{\bf O}&{\bf O}&{\bf O}&{\bf X}&{\bf O}&{\bf O}&
       &${\cal TR}$\\
$L_x$& &{\bf O}& & &{\bf O}&{\bf O}& &{\bf X}& &
       &{\bf O}&{\bf O}&{\bf O}&{\bf O}&{\bf O}& &{\bf X}& &
       &{\bf O}&{\bf O}&{\bf O}&{\bf O}&{\bf O}&{\bf O}&{\bf X}&{\bf O}& & \\
$L_z$& & &{\bf O}&{\bf O}& & &{\bf O}& &{\bf X}&
       &{\bf O}&{\bf O}&{\bf O}&{\bf O}& &{\bf O}& &{\bf X}&
       &{\bf O}&{\bf O}&{\bf O}&{\bf O}&{\bf O}&{\bf O}&{\bf O}&{\bf X}& & \\
 & & & & & & & & & & & & & & & & & & & & & & & & & & & & &
\end{tabular}
\end{table}

\begin{table}
\squeezetable
\caption{}
\label{tbl:num3}
\vspace*{4mm}
\begin{tabular}{c*{28}{c}c}
 & & & & & & & & & & & & & & & & & & & & & & & & & & & & &  \\
$G$  & &{\bf X}& &{\bf O}& &{\bf O}&{\bf O}&{\bf O}&{\bf O}&
       &{\bf X}&{\bf O}&{\bf O}&{\bf O}&{\bf O}&{\bf O}&{\bf O}&{\bf O}&
       &{\bf X}& &{\bf O}& & &{\bf O}&{\bf O}& & &  \\
$H$  & & &{\bf X}& &{\bf O}&{\bf O}&{\bf O}&{\bf O}&{\bf O}&
       &{\bf O}&{\bf X}&{\bf O}&{\bf O}&{\bf O}&{\bf O}&{\bf O}&{\bf O}&
       & &{\bf X}& &{\bf O}&{\bf O}& & &{\bf O}& &${\cal BT}$\\
$E$  & &{\bf O}& &{\bf X}& &{\bf O}&{\bf O}&{\bf O}&{\bf O}&
       &{\bf O}&{\bf O}&{\bf X}&{\bf O}&{\bf O}&{\bf O}&{\bf O}&{\bf O}&
       &{\bf O}& &{\bf X}& &{\bf O}& & &{\bf O}& &  \\
$F$  & & &{\bf O}& &{\bf X}&{\bf O}&{\bf O}&{\bf O}&{\bf O}&
       &{\bf O}&{\bf O}&{\bf O}&{\bf X}&{\bf O}&{\bf O}&{\bf O}&{\bf O}&
       & &{\bf O}& &{\bf X}& &{\bf O}&{\bf O}& & &  \\
 & & & & & & & & & & & & & & & & & & & & & & & & & & & & &  \\
$O_x$& &{\bf X}&{\bf X}&{\bf X}&{\bf X}&{\bf X}&{\bf X}&{\bf X}&{\bf X}&
       &{\bf X}&{\bf X}&{\bf X}&{\bf X}&{\bf X}&{\bf X}&{\bf X}&{\bf X}&
       &{\bf X}&{\bf X}&{\bf X}&{\bf X}&{\bf X}&{\bf X}&{\bf X}&{\bf X}& & \\
$O_z$& &{\bf X}&{\bf X}&{\bf X}&{\bf X}&{\bf X}&{\bf X}&{\bf X}&{\bf X}&
       & & & & & & & & & & & & & & & & & & &${\cal BR}$\\
$C_x$& & & & & & & & & &
       &{\bf X}&{\bf X}&{\bf X}&{\bf X}&{\bf X}&{\bf X}&{\bf X}&{\bf X}&
       & & & & & & & & & &  \\
$C_z$& & & & & & & & & & & & & & & & & & &
       &{\bf X}&{\bf X}&{\bf X}&{\bf X}&{\bf X}&{\bf X}&{\bf X}&{\bf X}& & \\
 & & & & & & & & & & & & & & & & & & & & & & & & & & & & &  \\
$T_x$& &{\bf O}&{\bf O}&{\bf O}&{\bf O}&{\bf X}&{\bf O}&{\bf O}&{\bf O}&
       &{\bf O}&{\bf O}&{\bf O}&{\bf O}&{\bf X}&{\bf O}& & &
       & &{\bf O}&{\bf O}& &{\bf X}&{\bf O}& & & &  \\
$T_z$& &{\bf O}&{\bf O}&{\bf O}&{\bf O}&{\bf O}&{\bf X}&{\bf O}&{\bf O}&
       &{\bf O}&{\bf O}&{\bf O}&{\bf O}&{\bf O}&{\bf X}& & &
       &{\bf O}& & &{\bf O}&{\bf O}&{\bf X}& & & &${\cal TR}$\\
$L_x$& &{\bf O}&{\bf O}&{\bf O}&{\bf O}&{\bf O}&{\bf O}&{\bf X}&{\bf O}&
       &{\bf O}&{\bf O}&{\bf O}&{\bf O}& & &{\bf X}&{\bf O}&
       &{\bf O}& & &{\bf O}& & &{\bf X}&{\bf O}& &  \\
$L_z$& &{\bf O}&{\bf O}&{\bf O}&{\bf O}&{\bf O}&{\bf O}&{\bf O}&{\bf X}&
       &{\bf O}&{\bf O}&{\bf O}&{\bf O}& & &{\bf O}&{\bf X}&
       & &{\bf O}&{\bf O}& & & &{\bf O}&{\bf X}& &  \\
 & & & & & & & & & & & & & & & & & & & & & & & & & & & & &
\end{tabular}
\end{table}

\begin{table}
\squeezetable
\caption{}
\label{tbl:num4}
\vspace*{4mm}
\begin{tabular}{c*{28}{c}c}
 & & & & & & & & & & & & & & & & & & & & & & & & & & & & &  \\
$G$  & &{\bf X}& &{\bf O}& &{\bf O}& & &{\bf O}&
       &{\bf X}&{\bf O}&{\bf O}&{\bf O}&{\bf O}&{\bf O}&{\bf O}&{\bf O}&
       &{\bf X}& &{\bf O}& &{\bf O}&{\bf O}&{\bf O}&{\bf O}& &  \\
$H$  & & &{\bf X}& &{\bf O}& &{\bf O}&{\bf O}& &
       &{\bf O}&{\bf X}&{\bf O}&{\bf O}&{\bf O}&{\bf O}&{\bf O}&{\bf O}&
       & &{\bf X}& &{\bf O}&{\bf O}&{\bf O}&{\bf O}&{\bf O}& &${\cal BT}$\\
$E$  & &{\bf O}& &{\bf X}& & &{\bf O}&{\bf O}& &
       &{\bf O}&{\bf O}&{\bf X}&{\bf O}&{\bf O}&{\bf O}&{\bf O}&{\bf O}&
       &{\bf O}& &{\bf X}& &{\bf O}&{\bf O}&{\bf O}&{\bf O}& &  \\
$F$  & & &{\bf O}& &{\bf X}&{\bf O}& & &{\bf O}&
       &{\bf O}&{\bf O}&{\bf O}&{\bf X}&{\bf O}&{\bf O}&{\bf O}&{\bf O}&
       & &{\bf O}& &{\bf X}&{\bf O}&{\bf O}&{\bf O}&{\bf O}& &  \\
 & & & & & & & & & & & & & & & & & & & & & & & & & & & & &  \\
$O_x$& & & & & & & & & & & & & & & & & & & & & & & & & & & & &  \\
$O_z$& &{\bf X}&{\bf X}&{\bf X}&{\bf X}&{\bf X}&{\bf X}&{\bf X}&{\bf X}&
       &{\bf X}&{\bf X}&{\bf X}&{\bf X}&{\bf X}&{\bf X}&{\bf X}&{\bf X}&
       & & & & & & & & & &${\cal BR}$\\
$C_x$& &{\bf X}&{\bf X}&{\bf X}&{\bf X}&{\bf X}&{\bf X}&{\bf X}&{\bf X}&
       & & & & & & & & &
       &{\bf X}&{\bf X}&{\bf X}&{\bf X}&{\bf X}&{\bf X}&{\bf X}&{\bf X}& & \\
$C_z$& & & & & & & & & &
       &{\bf X}&{\bf X}&{\bf X}&{\bf X}&{\bf X}&{\bf X}&{\bf X}&{\bf X}&
       &{\bf X}&{\bf X}&{\bf X}&{\bf X}&{\bf X}&{\bf X}&{\bf X}&{\bf X}& & \\
 & & & & & & & & & & & & & & & & & & & & & & & & & & & & &  \\
$T_x$& &{\bf O}& & &{\bf O}&{\bf X}&{\bf O}& & &
       &{\bf O}&{\bf O}&{\bf O}&{\bf O}&{\bf X}&{\bf O}& & &
       &{\bf O}&{\bf O}&{\bf O}&{\bf O}&{\bf X}&{\bf O}&{\bf O}&{\bf O}& & \\
$T_z$& & &{\bf O}&{\bf O}& &{\bf O}&{\bf X}& & &
       &{\bf O}&{\bf O}&{\bf O}&{\bf O}&{\bf O}&{\bf X}& & &
       &{\bf O}&{\bf O}&{\bf O}&{\bf O}&{\bf O}&{\bf X}&{\bf O}&{\bf O}&
       &${\cal TR}$\\
$L_x$& & &{\bf O}&{\bf O}& & & &{\bf X}&{\bf O}&
       &{\bf O}&{\bf O}&{\bf O}&{\bf O}& & &{\bf X}&{\bf O}&
       &{\bf O}&{\bf O}&{\bf O}&{\bf O}&{\bf O}&{\bf O}&{\bf X}&{\bf O}& & \\
$L_z$& &{\bf O}& & &{\bf O}& & &{\bf O}&{\bf X}&
       &{\bf O}&{\bf O}&{\bf O}&{\bf O}& & &{\bf O}&{\bf X}&
       &{\bf O}&{\bf O}&{\bf O}&{\bf O}&{\bf O}&{\bf O}&{\bf O}&{\bf X}& & \\
 & & & & & & & & & & & & & & & & & & & & & & & & & & & & &
\end{tabular}
\end{table}

\begin{table}
\squeezetable
\caption{}
\label{tbl:num5}
\vspace*{4mm}
\begin{tabular}{c*{28}{c}c}
 & & & & & & & & & & & & & & & & & & & & & & & & & & & & &  \\
$G$  & &{\bf X}&{\bf O}& & &{\bf O}&{\bf O}&{\bf O}&{\bf O}&
       &{\bf X}&{\bf O}&{\bf O}&{\bf O}&{\bf O}&{\bf O}&{\bf O}&{\bf O}&
       &{\bf X}&{\bf O}& & & &{\bf O}&{\bf O}& & &  \\
$H$  & &{\bf O}&{\bf X}& & &{\bf O}&{\bf O}&{\bf O}&{\bf O}&
       &{\bf O}&{\bf X}&{\bf O}&{\bf O}&{\bf O}&{\bf O}&{\bf O}&{\bf O}&
       &{\bf O}&{\bf X}& & &{\bf O}& & &{\bf O}& &${\cal BT}$\\
$E$  & & & &{\bf X}&{\bf O}&{\bf O}&{\bf O}&{\bf O}&{\bf O}&
       &{\bf O}&{\bf O}&{\bf X}&{\bf O}&{\bf O}&{\bf O}&{\bf O}&{\bf O}&
       & & &{\bf X}&{\bf O}&{\bf O}& & &{\bf O}& &  \\
$F$  & & & &{\bf O}&{\bf X}&{\bf O}&{\bf O}&{\bf O}&{\bf O}&
       &{\bf O}&{\bf O}&{\bf O}&{\bf X}&{\bf O}&{\bf O}&{\bf O}&{\bf O}&
       & & &{\bf O}&{\bf X}& &{\bf O}&{\bf O}& & &  \\
 & & & & & & & & & & & & & & & & & & & & & & & & & & & & &  \\
$O_x$& &{\bf O}&{\bf O}&{\bf O}&{\bf O}&{\bf X}&{\bf O}&{\bf O}&{\bf O}&
       &{\bf O}&{\bf O}&{\bf O}&{\bf O}&{\bf X}&{\bf O}& & &
       & &{\bf O}&{\bf O}& &{\bf X}&{\bf O}& & & &  \\
$O_z$& &{\bf O}&{\bf O}&{\bf O}&{\bf O}&{\bf O}&{\bf X}&{\bf O}&{\bf O}&
       &{\bf O}&{\bf O}&{\bf O}&{\bf O}&{\bf O}&{\bf X}& & &
       &{\bf O}& & &{\bf O}&{\bf O}&{\bf X}& & & &${\cal BR}$\\
$C_x$& &{\bf O}&{\bf O}&{\bf O}&{\bf O}&{\bf O}&{\bf O}&{\bf X}&{\bf O}&
       &{\bf O}&{\bf O}&{\bf O}&{\bf O}& & &{\bf X}&{\bf O}&
       &{\bf O}& & &{\bf O}& & &{\bf X}&{\bf O}& &  \\
$C_z$& &{\bf O}&{\bf O}&{\bf O}&{\bf O}&{\bf O}&{\bf O}&{\bf O}&{\bf X}&
       &{\bf O}&{\bf O}&{\bf O}&{\bf O}& & &{\bf O}&{\bf X}&
       & &{\bf O}&{\bf O}& & & &{\bf O}&{\bf X}& &  \\
 & & & & & & & & & & & & & & & & & & & & & & & & & & & & &  \\
$T_x$& &{\bf X}&{\bf X}&{\bf X}&{\bf X}&{\bf X}&{\bf X}&{\bf X}&{\bf X}&
       &{\bf X}&{\bf X}&{\bf X}&{\bf X}&{\bf X}&{\bf X}&{\bf X}&{\bf X}&
       &{\bf X}&{\bf X}&{\bf X}&{\bf X}&{\bf X}&{\bf X}&{\bf X}&{\bf X}& & \\
$T_z$& &{\bf X}&{\bf X}&{\bf X}&{\bf X}&{\bf X}&{\bf X}&{\bf X}&{\bf X}&
       & & & & & & & & & & & & & & & & & & &${\cal TR}$\\
$L_x$& & & & & & & & & &
       &{\bf X}&{\bf X}&{\bf X}&{\bf X}&{\bf X}&{\bf X}&{\bf X}&{\bf X}&
       & & & & & & & & & &  \\
$L_z$& & & & & & & & & & & & & & & & & & &
       &{\bf X}&{\bf X}&{\bf X}&{\bf X}&{\bf X}&{\bf X}&{\bf X}&{\bf X}& & \\
 & & & & & & & & & & & & & & & & & & & & & & & & & & & & &
\end{tabular}
\end{table}

\begin{table}
\squeezetable
\caption{}
\label{tbl:num6}
\vspace*{4mm}
\begin{tabular}{c*{28}{c}c}
 & & & & & & & & & & & & & & & & & & & & & & & & & & & & &  \\
$G$  & &{\bf X}&{\bf O}& & &{\bf O}& & &{\bf O}&
       &{\bf X}&{\bf O}&{\bf O}&{\bf O}&{\bf O}&{\bf O}&{\bf O}&{\bf O}&
       &{\bf X}&{\bf O}& & &{\bf O}&{\bf O}&{\bf O}&{\bf O}& &  \\
$H$  & &{\bf O}&{\bf X}& & & &{\bf O}&{\bf O}& &
       &{\bf O}&{\bf X}&{\bf O}&{\bf O}&{\bf O}&{\bf O}&{\bf O}&{\bf O}&
       &{\bf O}&{\bf X}& & &{\bf O}&{\bf O}&{\bf O}&{\bf O}& &${\cal BT}$\\
$E$  & & & &{\bf X}&{\bf O}& &{\bf O}&{\bf O}& &
       &{\bf O}&{\bf O}&{\bf X}&{\bf O}&{\bf O}&{\bf O}&{\bf O}&{\bf O}&
       & & &{\bf X}&{\bf O}&{\bf O}&{\bf O}&{\bf O}&{\bf O}& &  \\
$F$  & & & &{\bf O}&{\bf X}&{\bf O}& & &{\bf O}&
       &{\bf O}&{\bf O}&{\bf O}&{\bf X}&{\bf O}&{\bf O}&{\bf O}&{\bf O}&
       & & &{\bf O}&{\bf X}&{\bf O}&{\bf O}&{\bf O}&{\bf O}& &  \\
 & & & & & & & & & & & & & & & & & & & & & & & & & & & & &  \\
$O_x$& &{\bf O}& & &{\bf O}&{\bf X}&{\bf O}& & &
       &{\bf O}&{\bf O}&{\bf O}&{\bf O}&{\bf X}&{\bf O}& & &
       &{\bf O}&{\bf O}&{\bf O}&{\bf O}&{\bf X}&{\bf O}&{\bf O}&{\bf O}& & \\
$O_z$& & &{\bf O}&{\bf O}& &{\bf O}&{\bf X}& & &
       &{\bf O}&{\bf O}&{\bf O}&{\bf O}&{\bf O}&{\bf X}& & &
       &{\bf O}&{\bf O}&{\bf O}&{\bf O}&{\bf O}&{\bf X}&{\bf O}&{\bf O}&
       &${\cal BR}$\\
$C_x$& & &{\bf O}&{\bf O}& & & &{\bf X}&{\bf O}&
       &{\bf O}&{\bf O}&{\bf O}&{\bf O}& & &{\bf X}&{\bf O}&
       &{\bf O}&{\bf O}&{\bf O}&{\bf O}&{\bf O}&{\bf O}&{\bf X}&{\bf O}& & \\
$C_z$& &{\bf O}& & &{\bf O}& & &{\bf O}&{\bf X}&
       &{\bf O}&{\bf O}&{\bf O}&{\bf O}& & &{\bf O}&{\bf X}&
       &{\bf O}&{\bf O}&{\bf O}&{\bf O}&{\bf O}&{\bf O}&{\bf O}&{\bf X}& & \\
 & & & & & & & & & & & & & & & & & & & & & & & & & & & & &  \\
$T_x$& & & & & & & & & & & & & & & & & & & & & & & & & & & & &  \\
$T_z$& &{\bf X}&{\bf X}&{\bf X}&{\bf X}&{\bf X}&{\bf X}&{\bf X}&{\bf X}&
       &{\bf X}&{\bf X}&{\bf X}&{\bf X}&{\bf X}&{\bf X}&{\bf X}&{\bf X}&
       & & & & & & & & & &${\cal TR}$\\
$L_x$& &{\bf X}&{\bf X}&{\bf X}&{\bf X}&{\bf X}&{\bf X}&{\bf X}&{\bf X}&
       & & & & & & & & &
       &{\bf X}&{\bf X}&{\bf X}&{\bf X}&{\bf X}&{\bf X}&{\bf X}&{\bf X}& & \\
$L_z$& & & & & & & & & &
       &{\bf X}&{\bf X}&{\bf X}&{\bf X}&{\bf X}&{\bf X}&{\bf X}&{\bf X}&
       &{\bf X}&{\bf X}&{\bf X}&{\bf X}&{\bf X}&{\bf X}&{\bf X}&{\bf X}& & \\
 & & & & & & & & & & & & & & & & & & & & & & & & & & & & &
\end{tabular}
\end{table}


\begin{thebibliography}{99}
\bibitem[*]{byline}  Research supported in part by the U.~S.~
National Science Foundation.
\bibitem{BDS} I. S. Barker, A. Donnachie, and J. K. Storrow,
  {\it Nucl. Phys.} {\bf B95}, 347 (1975).
\bibitem{Workman} G. Keaton and R. Workman, {\it Phys. Rev.} {\bf C53},
  1434 (1996).
\bibitem{vmeson} M. Pichowsky, \c{C}. \c{S}avkli and F. Tabakin,
  {\it Nucl. Phys.} {\bf A370}, 311 (1994).
\bibitem{psmeson} C. G. Fasano, F. Tabakin, and B. Saghai,
  {\it Phys. Rev.} {\bf C46}, 2430 (1992).
\bibitem{Morav} M. J. Moravcsik, {\it Phys. Rev.} {\bf D29}, 2625 (1984);
  {\it J. Math. Phys.} {\bf 26}, 211 (1995).
\bibitem{DL} N. W. Dean and P. Lee, {\it Phys. Rev.} {\bf D5}, 2741 (1972).
\bibitem{Fierz} C. Itzykson and J.-B. Zuber, {\it Quantum Field Theory},
  (McGraw-Hill, New York, 1985), p. 161; M. E. Peskin and D. V. Schroeder,
  {\it An Introduction to Quantum Field Theory}, (Addison-Wesley, Reading,
  1995), p. 75.
\bibitem{Goldstein} G. R. Goldstein, J. F. Owens III, J. P Rutherfoord, and
  M. J. Moravcsik, {\it Nucl. Phys.} {\bf B80}, 164 (1974).


\end{thebibliography}
\end{document}